# Concurrent enforcement of polyconvexity and true-stress–true-strain monotonicity in incompressible isotropic hyperelasticity: application to neural network constitutive models

Maximilian P. Wollner [a,*], Dominik K. Klein [b], Herbert Baaser [c], Gerhard A. Holzapfel [a,d], Patrizio Neff [e]

[a] Institute of Biomechanics, Graz University of Technology, Stremayrgasse 16/2, Graz, 8010, Austria
[b] Cyber-Physical Simulation, Department of Mechanical Engineering, TU Darmstadt, 64293, Darmstadt, Germany
[c] University of Applied Sciences Bingen, Berlinstr. 109, 55411, Bingen, Germany
[d] Department of Structural Engineering, Norwegian University of Science and Technology, Høgskoleringen 6, Trondheim, 7491, Norway
[e] Chair of Nonlinear Analysis and Modelling, University of Duisburg-Essen, Thea-Leymann-Straße 9, Essen, 45127, Germany

ABSTRACT

The design of physics-augmented neural networks (PANNs) for the purposes of constitutive modeling has received considerable attention as of late for a variety of material behaviors. Here, we revisit the classical framework of isotropic incompressible hyperelasticity in light of recent advances in the study of constitutive inequalities. We show that polyconvexity implies true-stress–true-strain monotonicity for a large class of incompressible strain-energy functions. The resulting elastic law obeys the physically reasonable Legendre–Hadamard (or ellipticity) condition as well as the notion of increasing stress with increasing strain. These results then inform the architecture of four distinct PANNs which are subsequently calibrated to three different sets of experimental data each. We show that different PANN parametrizations – satisfying the same constitutive constraints *a priori* – have varying approximation power for the description of material behavior. Moreover, even when distinct parametrizations perform comparatively well within the calibration regime, they show pronounced differences in extrapolation. This observation motivates a critical discussion about the predictive power of PANNs which also has implications for the modeling of more complex material behavior by virtue of neural networks.

## 1. Introduction

The study of incompressible isochoric hyperelasticity has a long history dating back to the initial works by Mooney (1940), Wall (1942a,b), and Rivlin (1948).[1] Since then a great number of functional forms for strain-energy functions have been proposed for both invariant-based and principal stretch-based models, cf. Steinmann et al. (2012), Hossain and Steinmann (2013), and

* Corresponding author.
*E-mail address:* wollner@tugraz.at (M.P. Wollner).

[1] While there is experimental evidence that many rubbers and rubber-like materials deform approximately isochorically in certain boundary-value problems, cf. Holt and McPherson (1936), we should keep in mind that incompressibility is much more a statement about the ratio of 'volumetric stiffness' to 'deviatoric stiffness'; rubber can be compressed by sufficiently large hydrostatic pressures, cf. Adams and Gibson (1930). The assumption of incompressibility is also in large parts a simple mathematical convenience. From a historical perspective, one should not forget that many boundary-value problems in isotropic hyperelasticity only allow analytical solutions because of assumed incompressibility,

Ricker and Wriggers (2023). However, the choice and calibration of a suitable constitutive model out of the multitude available is cumbersome and requires a lot of expert knowledge, cf. Baaser (2026b), Baaser and Becker (2023). This issue can be overcome by virtue of constitutive models based on physics-augmented neural networks (PANNs).[2] These function approximators offer considerably more flexibility than aforementioned conventional modeling approaches and have been shown to be applicable to a wide range of different material behaviors. This largely facilitates the challenging process of selecting an appropriate constitutive model from the broad range available. Although the research on incompressible hyperelasticity and PANN constitutive modeling is already highly advanced, we nonetheless return to this admittedly simple framework for two principal reasons. First, recent advances concerning constitutive inequalities in isotropic hyperelasticity have again raised questions about the fringe case of incompressibility, cf. Wollner et al. (2026a) and Baaser (2026a). Second, the extrapolation of PANNs *far away* from the training data has received comparatively little attention. To this end, the case of isotropic incompressible hyperelasticity offers an illustrative testing ground, in particular in light of the aforementioned constitutive constraints.

In the context of compressible isotropic hyperelasticity, Linden et al. (2023, Sect. 2.2) lists a number of requirements that are widely considered physically reasonable: material frame-indifference, material symmetry, a stress-free and energy-free undeformed configuration, polyconvexity and in turn rank-one convexity, and volumetric growth conditions. In addition, the general expectation for an isotropic hyperelastic material is to exhibit increasing levels of stress with increasing levels of strain. As recently shown in Wollner et al. (2026a, Prop. 5.7), polyconvexity alone does not ensure a monotonic true-stress response in unconstrained uniaxial extension. It can therefore be argued that an additional *true-stress–true-strain monotonicity* (TSTS-M) should be added to the list of physically reasonable constitutive constraints, cf. Leblond (1992), d'Agostino et al. (2025), and Neff et al. (2025a). At the time of writing, the authors do not know of any compressible strain-energy function that satisfies polyconvexity and TSTS-M simultaneously for all admissible deformation states. In the highly restrictive case of incompressibility, however, the situation simplifies considerably. As demonstrated in this work, TSTS-M is automatically implied for a large class of polyconvex strain-energy functions in case of incompressibility. In particular, we show:

- a more traceable version of the original proof by Ball (1976) which provides sufficient conditions for polyconvexity by means of the theory of majorization,
- that Ball's sufficient conditions for polyconvexity imply TSTS-M in case of incompressibility,
- that Ball's sufficient conditions lead to straightforward requirements for polyconvexity and in turn TSTS-M in a large class of invariant-based parametrizations.

Ball's sufficient conditions for polyconvexity apply to a wide range of hyperelastic models in the literature. Still, it is important to note that these conditions are not necessary for polyconvexity and more flexible parameterizations have been proposed, e.g., cf. Wiedemann and Peter (2026). Consequently, our results do not imply that polyconvexity generally ensures TSTS-M in case of incompressibility.

Concerning our second point of interest, a constitutive model extrapolates in two qualitatively distinct ways: (i) from observed deformation modes to unknown ones and (ii) across some observed strain regime. A discussion of the first aspect can be found in Steinmann et al. (2012), Dammaß et al. (2025), and Klein et al. (2026a). The second type of extrapolation has received considerably less attention. Here, it is helpful to compare PANNs with classic approaches. While a limiting-chain model like Arruda and Boyce (1993), Gent (1996), or Pelliciari et al. (2026) always ensures that the stress diverges at a particular level of deformation, a hyperelastic PANN does generally not include such additional information. Even if TSTS-M is prescribed, the extrapolation of a PANN might not actually reflect our expectation of how a rubber-like material should behave. This is the primary focus of the second part of this contribution, where:

- we construct four distinct neural network architectures in line with the constitutive constraints set out in the first part,
- calibrate each to three sets of experimental data of distinct soft rubber-like materials,
- and compare their quality of interpolation and more importantly extrapolation.

We structure this work as follows: In Section 2, we review several constitutive constraints in hyperelasticity and discuss the implications of incompressibility, namely polyconvexity, TSTS-M, and Hill's inequality. We continue by revisiting the proof of Ball's conditions for polyconvexity and prove a number of implications for TSTS-M and invariant-based parametrizations in Section 3. These results are subsequently used in the construction of four different neural network architectures in Section 4 which are then calibrated and compared in Section 5. We finish this contribution with a short summary and discussion in Section 6.

## 2. Review of some constitutive constraints

The only constraint placed on a hyperelastic stress response by thermodynamics is its derivation from a strain-energy function $W(\mathbf{F})$, where $\mathbf{F}$ denotes the deformation gradient. Setting aside more obvious requirements such as material-frame indifference and cf. Truesdell and Noll (1965, Chap. D.Ib)) and Ogden (1997, Chap. 5). This is obviously not to say that the study of incompressible material behavior is not worthwhile from a physical and mathematical point of view.

[2] Related approaches are commonly referred to as thermodynamics-based, cf. Masi and Stefanou (2022), mechanics-informed, cf. As'ad and Farhat (2023), physics-based, cf. Aldakheel et al. (2025), physics-constrained, cf. Kalina et al. (2023), or constitutive artificial neural networks, cf. Linka et al. (2021). A variety of models have been proposed for hyperelasticity, cf. Fuhg et al. (2025, Sect. 4.2) for a review.

material symmetry, one can define additional constitutive constraints that limit the functional form of $W$ in order to satisfy certain desired properties *a priori*. This task has been dubbed Truesdell's Hauptproblem, cf. Truesdell (1956), and has attracted the attention of a number of authors, e.g., cf. Hill (1968, 1970), Krawietz (1975), Ball (1976, 1977), Leblond (1992), and Neff et al. (2025b) to name a few. An overview of the many constitutive inequalities in finite elasticity can be found in Truesdell and Noll (1965, Sects. 51–53), Šilhavý (1997, Sect. 18.6), Rivlin (2004), Neff et al. (2015), Mihai and Goriely (2017), and Wollner et al. (2026a). Here, we only want to revisit some established results to give the appropriate context for the novel findings that are presented in Section 3.

### 2.1. Polyconvexity

In Morrey (1952), the notion of *quasiconvexity* was used to establish existence results in the Calculus of Variations, which Ball (1976) subsequently applied to non-linear elasticity in his hallmark paper. Given the definition of quasiconvexity as an integral constraint on $W$, it is difficult to enforce *a priori*. Hence, Ball (1976) proposed the slightly stricter requirement of *polyconvexity* which implies quasiconvexity, but is easier to prescribe as a constitutive requirement, cf. Krawietz (1986, Sect. 12.4). The polyconvexity of $W$ is ensured if there exists a function $G(\mathbf{F}, \mathbf{A}, \delta)$, which is convex over $\mathbb{R}^{3\times3} \times \mathbb{R}^{3\times3} \times \mathbb{R}_+$, such that

$$W(\mathbf{F}) = G(\mathbf{F}, \operatorname{Cof}\mathbf{F}, \det\mathbf{F}), \tag{2.1}$$

cf. Ball (1976, Eq. (0.8)). In combination with suitable growth conditions on $W$, polyconvexity ensures the existence of minimizers in a hyperelastic boundary-value problem. More importantly, polyconvexity also has the advantage of implying rank-one convexity (or Legendre–Hadamard ellipticity), cf. Šilhavý (1997, Sect. 17.3), which guarantees stability and real-wave speeds to infinitesimal disturbances, cf. Truesdell and Noll (1965, Sect. 68 bis. and 71), and can otherwise be more difficult to establish, cf. Zee and Sternberg (1983, Sect. 3) and Aubert (1988, Thm. 4.2). From a constitutive modeling perspective, we employ polyconvexity for its implication of rank-one convexity, rather than for its significance in existence theorems. In case of incompressibility ($\det\mathbf{F} = 1$), polyconvexity can be ensured by some convex function $G(\mathbf{F}, \mathbf{A})$ with

$$W(\mathbf{F}) = G(\mathbf{F}, \operatorname{Cof}\mathbf{F}) = G(\mathbf{F}, \mathbf{F}^{-1}), \tag{2.2}$$

cf. Ball (1976, Sect. 8) and Ball (1977, Item (H1)'), where any dependence on the determinant simply drops out.[3,4] In conjunction with material frame-indifference and isotropy, Ball (1976, Thm. 5.2) established rather general sufficient conditions for polyconvexity by virtue of a representation in the principal stretches $\lambda_k$, which are the singular values of $\mathbf{F}$. Transitioning to a parametrization in terms of the signed singular values of $\mathbf{F}$ provides still additional leverage leading to not only sufficient, but also necessary conditions for polyconvexity, cf. Rosakis (1997), Mielke (2005), and recently Wiedemann and Peter (2026).

### 2.2. True-stress–true-strain monotonicity

Another line of constitutive inequalities in isotropic elasticity is related to the monotonicity of specific stress-strain pairs, cf. Ghiba et al. (2026). While it might seem intuitive that stress should increase with strain, the question in *which* stress measure and *which* strain measure such a condition is formulated is fundamental, cf. Neff et al. (2015). For instance, monotonicity constraints in the deformation gradient and the first Piola–Kirchhoff stress would imply convexity of the hyperelastic potential in the deformation gradient, which is not physically reasonable, cf. Ciarlet (1988, Sect. 4.8). Based on the pioneering work by Hill (1968, 1970), to which we will return in just a bit, Leblond (1992) took a closer look at the rate inequality

$$\left\langle \frac{\mathrm{D}^{\mathrm{ZJ}}\boldsymbol{\sigma}}{\mathrm{D}t}, \mathbf{D} \right\rangle = \langle \dot{\boldsymbol{\sigma}} + \boldsymbol{\sigma}\,\mathbf{W} - \mathbf{W}\,\boldsymbol{\sigma}, \mathbf{D} \rangle > 0 \qquad \forall\, \mathbf{D} \in \operatorname{Sym}(3) \setminus \{\mathbf{0}\}, \tag{2.3}$$

where $\frac{\mathrm{D}^{\mathrm{ZJ}}\boldsymbol{\sigma}}{\mathrm{D}t}$ denotes the Zaremba–Jaumann rate of the Cauchy (true) stress tensor $\boldsymbol{\sigma}$, while $\mathbf{D}$ and $\mathbf{W}$ are the symmetric and skew parts of the velocity gradient $\mathbf{L} = \dot{\mathbf{F}}\,\mathbf{F}^{-1}$, respectively.[5] Physically, we may interpret the above inequality as the requirement that any change in actual deformation, encapsulated by the strain rate tensor $\mathbf{D}$, is accompanied by a corresponding change in stress, represented by the objective stress rate $\frac{\mathrm{D}^{\mathrm{ZJ}}\boldsymbol{\sigma}}{\mathrm{D}t}$.

In case of isotropic hyperelasticity, Leblond (1992, Eq. (23)) showed that

$$\left\langle \frac{\mathrm{D}^{\mathrm{ZJ}}\boldsymbol{\sigma}}{\mathrm{D}t}, \mathbf{D} \right\rangle > 0 \quad \forall\, \mathbf{D} \in \operatorname{Sym}(3) \setminus \{\mathbf{0}\} \qquad \Longleftrightarrow \qquad \left\langle \mathrm{D}_{\log\mathbf{V}}\widehat{\boldsymbol{\sigma}}.\mathbf{H}, \mathbf{H} \right\rangle > 0 \quad \forall\, \mathbf{H} \in \operatorname{Sym}(3) \setminus \{\mathbf{0}\}, \tag{2.4}$$

where $\mathbf{V} = \sqrt{\mathbf{F}\,\mathbf{F}^{\mathrm{T}}}$ denotes the left stretch tensor and $\boldsymbol{\sigma} = \widehat{\boldsymbol{\sigma}}(\log\mathbf{V})$. The equivalence in (2.4) also holds in case the elastic material law is solely Cauchy-elastic, cf. d'Agostino et al. (2025, App. A.4.2) and more broadly (Yavari and Goriely, 2025). The tensor logarithm $\log\mathbf{V}$ can readily be identified with the Hencky (true) strain tensor which takes center stage in the discussion of (2.3) and has a number of special geometrical properties that make it stand out from other strain measures, cf. Neff et al. (2016). In the following, we will refer to the second statement in (2.4) as the *true-stress–true-strain monotonicity* (TSTS-M).

[3] In nearly incompressible hyperelasticity, it is common to formulate the potential $W$ as a function of the isochoric deformation gradient $\bar{\mathbf{F}} = J^{-1/3}\mathbf{F}$ instead of the full deformation gradient $\mathbf{F}$, which can be beneficial from a physical and numerical perspective, cf. Sansour (2008) and Bonet et al. (2015). For the results obtained in our work, the use of either $\mathbf{F}$ or $\bar{\mathbf{F}}$ makes no difference, and for simplicity, we chose the former.

[4] The derivation of rank-one convexity and its implication by polyconvexity in case of incompressible hyperelasticity requires some additional nuances, cf. Zee and Sternberg (1983, Sect. 1) and Klein et al. (2026a, Sect. 2.2).

[5] Regarding notation, $\langle(\bullet),(\bullet)\rangle$ denotes the inner product between two tensors of equal order. The dot in $(\bullet).(\bullet)$ denotes the double contraction of a fourth-order tensor with a second-order tensor resulting in a second-order tensor.

Since TSTS-M must hold for all instances in time, it is perhaps unsurprising that the inequality (2.3) is equivalent to the strict monotonicity statement

$$\left\langle \widehat{\sigma}(\log \mathbf{V}_2) - \widehat{\sigma}(\log \mathbf{V}_1), \log \mathbf{V}_2 - \log \mathbf{V}_1 \right\rangle > 0 \qquad \forall\, \mathbf{V}_1 \neq \mathbf{V}_2, \tag{2.5}$$

cf. Neff et al. (2015, Rem. 4.1) and Martin and Neff (2015). We may interpret this implied global monotonicity condition somewhat casually as the multiaxial analogue of the notion that the Cauchy stress must be increasing with increasing Hencky strain which also seems a reasonable requirement for a purely elastic material law.

One valid objection to TSTS-M is its supposed reliance on a particular choice of objective stress rate. As it turns out, the equivalence in (2.4) also holds for the corotational logarithmic stress rate, i.e.,

$$\left\langle \frac{\mathrm{D}^{\log}\boldsymbol{\sigma}}{\mathrm{D}t}, \mathbf{D} \right\rangle > 0 \qquad \Longleftrightarrow \qquad \left\langle \mathrm{D}_{\log \mathbf{V}} \widehat{\boldsymbol{\sigma}}.\mathbf{H}, \mathbf{H} \right\rangle > 0, \tag{2.6}$$

cf. Neff et al. (2025a, Prop. 4.10). One can therefore extend the definition of the rate inequality to a *corotational stability postulate* (CSP) and conjecture that

$$\left\langle \frac{\mathrm{D}^{\circ}\boldsymbol{\sigma}}{\mathrm{D}t}, \mathbf{D} \right\rangle > 0 \qquad \Longleftrightarrow \qquad \left\langle \mathrm{D}_{\log \mathbf{V}} \widehat{\boldsymbol{\sigma}}.\mathbf{H}, \mathbf{H} \right\rangle > 0. \tag{2.7}$$

Here, '∘' denotes a placeholder for any structure-preserving corotational rate, cf. Neff et al. (2025c), which would give further support to TSTS-M. A proof of this general equivalence is in the works, cf. Voss et al. (2026). Finally, the condition (2.5) can be used for the proof of existence results in Cauchy-elastic boundary-value problems via transformation to a rate-formulation, cf. Neff et al. (2026).

### 2.3. TSTS-M and Hill's inequality in case of incompressibility

In this work we are primarily concerned with incompressible hyperelasticity, where $\det \mathbf{F} = 1$ and the Kirchhoff stress tensor coincides with the Cauchy stress tensor, i.e., $\boldsymbol{\tau} = \boldsymbol{\sigma}$. Hence, the rate constraint (2.3) reduces to

$$\left\langle \frac{\mathrm{D}^{\mathrm{ZJ}}\boldsymbol{\tau}}{\mathrm{D}t}, \mathbf{D} \right\rangle > 0 \qquad \forall\, \operatorname{tr} \mathbf{D} = 0. \tag{2.8}$$

Without the additional constraint $\operatorname{tr} \mathbf{D} = 0$, the above rate inequality is identical to an earlier constitutive constraint, namely *Hill's inequality*, reading simply

$$\left\langle \frac{\mathrm{D}^{\mathrm{ZJ}}\boldsymbol{\tau}}{\mathrm{D}t}, \mathbf{D} \right\rangle > 0, \tag{2.9}$$

cf. Hill (1968, 1970). The relation has been analyzed by Hill for generally compressible material behavior. In contrast to the Cauchy stress, the Kirchhoff stress is conjugate to the Hencky strain in case of isotropic hyperelasticity such that

$$\boldsymbol{\tau} = \mathrm{D}_{\log \mathbf{V}} \widehat{W} \qquad \text{with} \qquad W(\mathbf{F}) = \widehat{W}(\log \lambda_1, \log \lambda_2, \log \lambda_3), \tag{2.10}$$

cf. Murnaghan (1941, p. 127) and Richter (1948, Eq. (3.8*)). Hence, analogously to (2.4), Hill identified that

$$\left\langle \frac{\mathrm{D}^{\mathrm{ZJ}}\boldsymbol{\tau}}{\mathrm{D}t}, \mathbf{D} \right\rangle > 0 \qquad \Longleftrightarrow \qquad \left\langle \mathrm{D}_{\log \mathbf{V}} \widehat{\boldsymbol{\tau}}.\mathbf{H}, \mathbf{H} \right\rangle > 0 \qquad \Longleftrightarrow \qquad \left\langle \mathrm{D}^2_{\log \mathbf{V}} \widehat{W}.\mathbf{H}, \mathbf{H} \right\rangle > 0, \tag{2.11}$$

i.e., in the general compressible case the rate inequality (2.9) is satisfied if and only if $\widehat{W}$ is strictly convex in the Hencky strain or, equivalently, in $\log \lambda_k$. In the incompressible case (2.8), this strict convexity is still sufficient, but no longer necessary due to the additional constraint $\operatorname{tr} \mathbf{D} = 0$, cf. Baaser (2026a, App. D.2) and Appendix A. In fact, here

$$\left\langle \frac{\mathrm{D}^{\mathrm{ZJ}}\boldsymbol{\tau}}{\mathrm{D}t}, \mathbf{D} \right\rangle > 0 \quad \forall\, \operatorname{tr} \mathbf{D} = 0 \qquad \Longleftrightarrow \qquad \left\langle \mathrm{D}^2_{\log \mathbf{V}} \widehat{W}.\mathbf{H}, \mathbf{H} \right\rangle > 0 \quad \forall\, \operatorname{tr} \mathbf{H} = 0, \tag{2.12}$$

which is equivalent to requiring strict convexity of

$$\widehat{W}^{\mathrm{inc}}_{\mathrm{red}}(\log \lambda_1, \log \lambda_2) := \widehat{W}\left(\log \lambda_1, \log \lambda_2, -\log \lambda_1 - \log \lambda_2\right) \tag{2.13}$$

in the two remaining principal logarithmic strains, cf. Baaser (2026a, App. D.2–D.3). In the following, we refer to this requirement as the *weak form* of Hill's inequality.

Analogous to (2.5), Hill's inequality is equivalent to the global monotonicity statement

$$\left\langle \widehat{\boldsymbol{\tau}}(\log \mathbf{V}_2) - \widehat{\boldsymbol{\tau}}(\log \mathbf{V}_1), \log \mathbf{V}_2 - \log \mathbf{V}_1 \right\rangle > 0 \qquad \forall\, \mathbf{V}_1 \neq \mathbf{V}_2. \tag{2.14}$$

This is especially relevant in case of incompressibility, in which case it is in fact equivalent to (2.5) since $\boldsymbol{\tau} = \boldsymbol{\sigma}$. In contrast, in the general compressible case, Hill's inequality does not ensure monotonicity in the true-stress response, cf. Wollner et al. (2026a, Rem. 5.9). It is therefore only the particular case of incompressibility for which Hill's inequality is of interest.[6] An overview of the various constitutive constraints and their relations can be found in Fig. 1.

[6] Although Hill (1968, p. 239) touches on the constraint of incompressibility, the equivalence of (2.8) requires some additional attention. To the knowledge of the authors, this has not been discussed in the literature in closer detail. Hence, in the spirit of traceability and completeness, we have added Appendix A.

## 3. Combination of polyconvexity and TSTS-M

In the general compressible case, polyconvexity does not imply TSTS-M and *vice-versa*, which has already been observed by Leblond (1992, Sect. 4b). This result has recently also been highlighted by way of explicit examples in Wollner et al. (2026a). There, a polyconvex strain-energy is shown to produce a non-monotonic true-stress response in unconstrained uniaxial extension, while another TSTS-M conforming strain-energy function leads to a non-monotonic true-shear-stress response in simple shear. In fact, the authors do not know of a compressible strain-energy function that satisfies both polyconvexity and TSTS-M simultaneously for all deformation states.[7]

In the rather restricted case of incompressibility, the situation is fundamentally different. There exists a large number of strain-energy functions that satisfy both constitutive constraints at the same time, e.g., the Ogden model and in turn all its derivatives such as the Mooney–Rivlin model or neo-Hooke model. More generally, we can establish a new theoretical result proving that any incompressible strain-energy function that satisfies the sufficient condition for polyconvexity by Ball (1976) automatically ensures TSTS-M. This conclusion then offers several economic parametrizations for the construction of PANNs in the context of incompressible hyperelasticity.

### 3.1. Parametrization based on principal stretches

The principal goal of this section is to show that the sufficient conditions for polyconvexity by Ball (1976) are also sufficient for Hill's inequality and therefore TSTS-M in case of incompressibility. As defined shortly, these constraints are given in terms of the principal stretches, i.e., singular values of $\mathbf{F}$, which then straightforwardly imply polyconvexity and Hill's inequality for parametrizations of the strain-energy function $W$ in terms of invariants.

#### 3.1.1. Sufficient conditions for polyconvexity by virtue of Ball's theorem

Before discussing Hill's inequality, we need to revisit a theorem by Ball (1976, Thm. 5.2) concerning polyconvexity of isotropic strain-energy functions. In the following, we restate Ball's original result and then offer an extended explanation of the proof via an excursion to the theory of majorization, which we believe to be of value to the mechanics community.

We start out by introducing a function $g$ such that

$$G(\mathbf{F},\mathbf{A}) = g(\lambda_1,\lambda_2,\lambda_3,a_1,a_2,a_3) = g(\boldsymbol{\lambda},\boldsymbol{a}), \tag{3.1}$$

where $\boldsymbol{\lambda} = (\lambda_k)_{k=1}^3$ and $\boldsymbol{a} = (a_k)_{k=1}^3$ are the singular values of $\mathbf{F}$ and $\mathbf{A}$, respectively. By definition, we require the function $g$ to obey a permutation invariance of its arguments, namely it should remain unchanged under independent reordering of $\lambda_k$ and $a_k$, e.g., $g(\lambda_1,\lambda_2,\lambda_3,a_1,a_2,a_3) = g(\lambda_2,\lambda_1,\lambda_3,a_3,a_2,a_1)$.

With (2.2) and $\mathbf{A} = \operatorname{Cof}\mathbf{F}$, we have

$$W(\mathbf{F}) = g(\lambda_1,\lambda_2,\lambda_3,\lambda_2\lambda_3,\lambda_3\lambda_1,\lambda_1\lambda_2). \tag{3.2}$$

Given this association and the permutation invariance of $g$, the expression above remains unchanged for any reordering of $\lambda_k$, e.g., $g(\lambda_1,\lambda_2,\lambda_3,\lambda_2\lambda_3,\lambda_3\lambda_1,\lambda_1\lambda_2) = g(\lambda_2,\lambda_1,\lambda_3,\lambda_1\lambda_3,\lambda_3\lambda_2,\lambda_2\lambda_1)$.[8] Importantly, the inverse of $g$ is non-unique given some strain-energy function $W$. For example, $W(\mathbf{F}) = \operatorname{tr}\operatorname{Cof}\mathbf{V}$ can be expressed both by

$$g(\boldsymbol{\lambda},\boldsymbol{a}) = \lambda_1\lambda_2 + \lambda_2\lambda_3 + \lambda_3\lambda_1 \qquad \text{or} \qquad g(\boldsymbol{\lambda},\boldsymbol{a}) = a_1 + a_2 + a_3. \tag{3.3}$$

With the parametrization (3.1), we should be able to formulate conditions for $g$ so as to ensure the convexity of $G$ resulting in the (slightly adapted) theorem by Ball (1976, Thm. 5.2) in case of incompressibility:

**Theorem 3.1.** *(Ball (1976, Thm. 5.2)) Let $W$ be defined according to (3.2) and let $g$ be a convex and monotonically increasing function with permutation-invariance in its first three and last three arguments, respectively. Then $W$ is polyconvex.*

**Proof.** Let $\mathbf{F}_1,\mathbf{F}_2,\mathbf{A}_1,\mathbf{A}_2 \in \mathbb{R}^{3\times 3}$ alongside $\mathbf{F} = t\,\mathbf{F}_1 + (1-t)\,\mathbf{F}_2$ and $\mathbf{A} = t\,\mathbf{A}_1 + (1-t)\,\mathbf{A}_2$ $\forall t \in [0,1]$. Let the associated singular values be denoted by $\boldsymbol{\lambda}_1,\boldsymbol{\lambda}_2,\boldsymbol{a}_1,\boldsymbol{a}_2,\boldsymbol{\lambda},\boldsymbol{a} \in \mathbb{R}^3_+$, respectively.

Then, the properties of $g$ and definition (3.1) allow the following chain of implications:

$$\begin{aligned} G(\mathbf{F},\mathbf{A}) &= g(\boldsymbol{\lambda},\boldsymbol{a}) \\ &\le g(t\,\boldsymbol{\lambda}_1 + (1-t)\,\boldsymbol{\lambda}_2,\boldsymbol{a}) \\ &\le g(t\,\boldsymbol{\lambda}_1 + (1-t)\,\boldsymbol{\lambda}_2, t\,\boldsymbol{a}_1 + (1-t)\,\boldsymbol{a}_2) \\ &\le t\,g(\boldsymbol{\lambda}_1,\boldsymbol{a}_1) + (1-t)\,g(\boldsymbol{\lambda}_2,\boldsymbol{a}_2) \\ &= t\,G(\mathbf{F}_1,\mathbf{A}_1) + (1-t)\,G(\mathbf{F}_2,\mathbf{A}_2), \end{aligned} \tag{3.4}$$

i.e., the function $G$ is convex in its arguments which by definition (2.2) ensures that $W$ is polyconvex. □

---

7 For a valid candidate of such a strain-energy function, Patrizio Neff is offering a prize money of 500€, cf. Neff et al. (2025b). (Added in proof: Shortly before final publication, roughly at the same time, we received two independent solutions from Tongji University and Stanford University.)

8 The permutation-invariance in the arguments of $g$ must hold independently for $\boldsymbol{\lambda}$ and $\boldsymbol{a}$ which constitutes a stronger requirement than permutation invariance in $\boldsymbol{\lambda}$ after setting $\mathbf{A} = \operatorname{Cof}\mathbf{F}$. While this independence is needed for the original proof by Ball (1976), it can in fact be dropped, cf. Geuken et al. (2026, Sect. 3.3). We also do so for the construction of the corresponding PANN in Section 4.

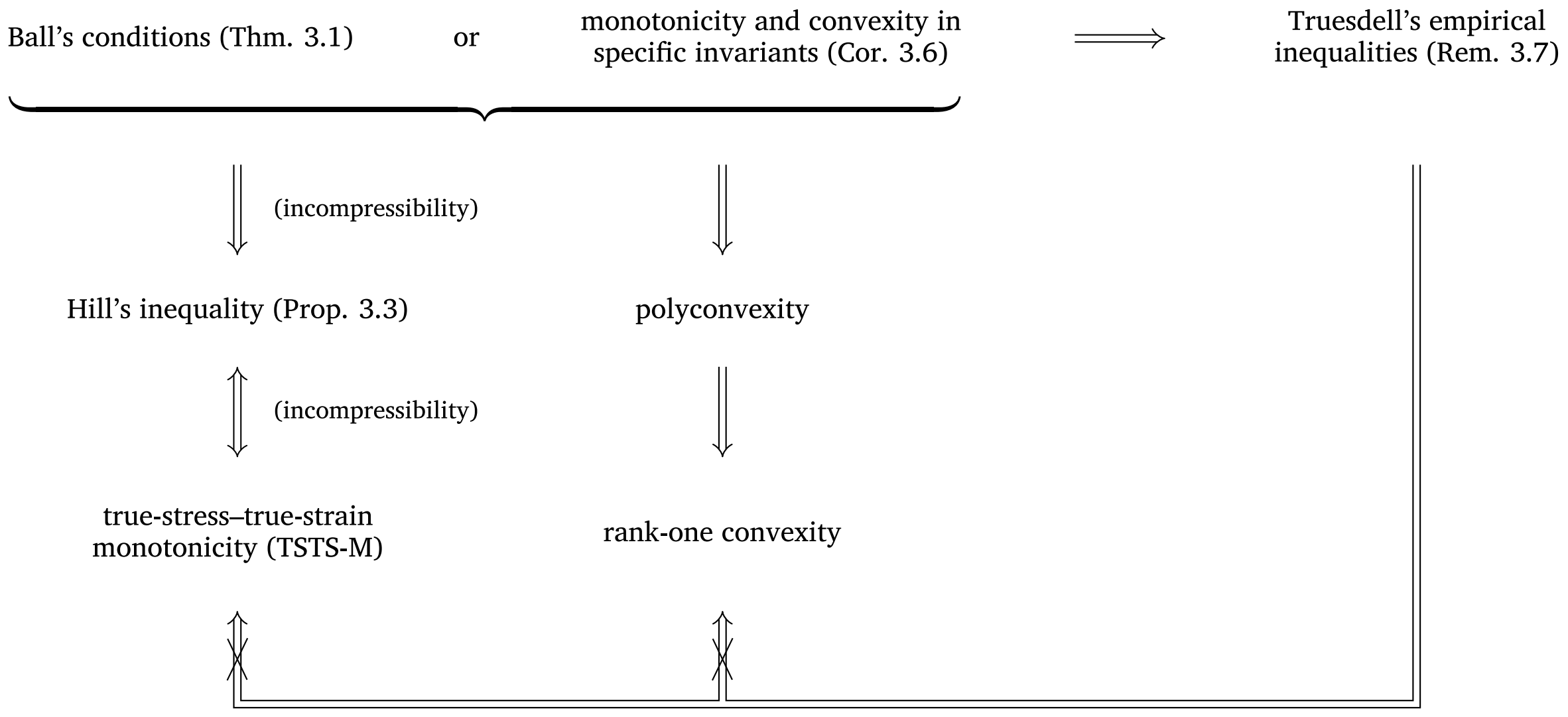

**Fig. 1.** Overview of various constitutive constraints and their relation in isotropic incompressible hyperelasticity.

While line four in (3.4) follows from the convexity of $g$ and line five is simply using definition (3.1), the legitimacy of the estimates in line two and three, i.e.,

$$g(\lambda, \boldsymbol{a}) \leq g(t\,\lambda_1 + (1-t)\,\lambda_2, \boldsymbol{a}) \leq g(t\,\lambda_1 + (1-t)\,\lambda_2, t\,\boldsymbol{a}_1 + (1-t)\,\boldsymbol{a}_2), \tag{3.5}$$

are not as straightforward to see. Initially one might suspect that these steps are the result of the monotonicity of $g$ in combination with supposed convexity of the ordered singular values in their associated tensor. As it turns out said convexity property is not given and an approach along such lines fails.[9] The result can instead be proven via the theory of majorization. The original proof by Ball (1976) seems to essentially make use of this concept, but never mentions it explicitly.

### 3.1.2. *Excursion into the theory of majorization*

Here, we want to take a short detour into the theory of majorization which might not be known to a larger part of the mechanics community, cf. also Lankeit et al. (2014). In the following, we will restrict ourselves to the three-dimensional case.

Take two sequence of positive real numbers in decreasing order, e.g., the singular values $\lambda_1 \geq \lambda_2 \geq \lambda_3$ and $\lambda'_1 \geq \lambda'_2 \geq \lambda'_3$, where

$$\lambda_1 \leq \lambda'_1, \tag{3.6a}$$

$$\lambda_1 + \lambda_2 \leq \lambda'_1 + \lambda'_2, \tag{3.6b}$$

$$\lambda_1 + \lambda_2 + \lambda_3 \leq \lambda'_1 + \lambda'_2 + \lambda'_3. \tag{3.6c}$$

Then $\lambda$ is said to be *weakly submajorized* by $\lambda'$, cf. Marshall et al. (2011, Def. A.2). Importantly, this does not necessitate that $\lambda_2 \leq \lambda'_2$ or $\lambda_3 \leq \lambda'_3$.

Interestingly, these inequalities imply a linear relationship between $\lambda$ and $\lambda'$ such that

$$\lambda = \boldsymbol{\mathcal{S}}\lambda', \tag{3.7}$$

where $\boldsymbol{\mathcal{S}} \in \mathbb{R}^{3\times 3}$ denotes a *doubly substochastic matrix* defined by

$$\mathcal{S}_{kl} \geq 0 \quad \text{and} \quad \sum_{k=1}^{3} \mathcal{S}_{kl} = \sum_{l=1}^{3} \mathcal{S}_{kl} \leq 1. \tag{3.8}$$

For a proof of this claim, the reader is referred to Marshall et al. (2011, Thm. C.4). A doubly substochastic matrix $\boldsymbol{\mathcal{S}}$ can be bounded by a *doubly stochastic matrix* $\boldsymbol{\mathcal{D}} \in \mathbb{R}^{3\times 3}$ defined by

$$\mathcal{D}_{kl} \geq 0 \quad \text{and} \quad \sum_{k=1}^{3} \mathcal{D}_{kl} = \sum_{l=1}^{3} \mathcal{D}_{kl} = 1. \tag{3.9}$$

[9] This can easily be seen by a counterexample: Take $\mathbf{F} = (\mathbf{F}_1 + \mathbf{F}_2)/2$ with $[\mathbf{F}_1] = \mathrm{diag}(3,2,1)$ and $[\mathbf{F}_2] = \mathrm{diag}(3,1,2)$ such that

$$\lambda_3(\mathbf{F}) = \frac{3}{2} > \frac{\lambda_3(\mathbf{F}_1) + \lambda_3(\mathbf{F}_2)}{2} = 1,$$

which clearly violates convexity.

Each entry of $\boldsymbol{\mathcal{S}}$ is bounded from above by $\boldsymbol{\mathcal{D}}$, i.e., $\mathcal{S}_{kl} \le \mathcal{D}_{kl}$. Since $\lambda' \in \mathbb{R}^3_+$, it immediately follows that

$$\lambda_k = \mathcal{S}_{kl}\,\lambda'_l \le \mathcal{D}_{kl}\,\lambda'_l. \tag{3.10}$$

There exists another useful feature of doubly stochastic matrices, referred to as Birkhoff's theorem, cf. Birkhoff (1946). Namely, any doubly stochastic matrix $\boldsymbol{\mathcal{D}}$ can be linearly decomposed into a series of *permutation matrices* $\boldsymbol{\mathcal{P}}_\alpha$, which are binary square matrices that rearrange the entries of any vector upon multiplication. Mathematically,

$$\boldsymbol{\mathcal{D}} = \sum_\alpha \omega_\alpha \boldsymbol{\mathcal{P}}_\alpha \qquad \text{with} \qquad \omega_\alpha \ge 0 \qquad \text{and} \qquad \sum_\alpha \omega_\alpha = 1. \tag{3.11}$$

A rather straightforward proof can also be found in Marshall et al. (2011, pp. 50–51).

### 3.1.3. *Application of majorization to isotropic hyperelasticity*

Having recalled some implications of the theory of majorization, we can establish the following result:

**Proposition 3.2.** *Let $f(\lambda)$ be a convex, monotonically increasing function, defined over $\lambda \in \mathbb{R}^3_+$, that is invariant under permutation of its arguments. Then,*

$$f(\lambda) \le f(t\,\lambda_1 + (1-t)\,\lambda_2), \tag{3.12}$$

*where $\lambda, \lambda_1, \lambda_2 \in \mathbb{R}^3_+$ are the singular values of $\mathbf{F} = t\,\mathbf{F}_1 + (1-t)\,\mathbf{F}_2$ $\forall t \in [0,1]$ and $\mathbf{F}_1, \mathbf{F}_2 \in \mathbb{R}^{3\times3}$, respectively.*

**Proof.** Following Ball (1976, Lems. 5.1–5.3), it is straightforward to establish that the sum of the first $n$ singular values is a convex function in the associated tensor, i.e.,

$$\sum_{k=1}^n \lambda_k(\mathbf{F}) = \sum_{k=1}^n \lambda_k(t\,\mathbf{F}_1 + (1-t)\,\mathbf{F}_2) \le \sum_{k=1}^n t\,\lambda_k(\mathbf{F}_1) + (1-t)\,\lambda_k(\mathbf{F}_2) \quad \forall n \in \{1,2,3\}. \tag{3.13}$$

Defining $\lambda' = t\,\lambda_1 + (1-t)\,\lambda_2$ and comparing (3.13) with (3.6), we see that $\lambda$ is weakly submajorized by $\lambda'$.

By virtue of (3.7), we then have

$$f(\lambda) = f(\boldsymbol{\mathcal{S}}\lambda') \tag{3.14}$$

for some doubly substochastic matrix $\boldsymbol{\mathcal{S}}$. Since the function $f$ is monotonically increasing, it follows with (3.10) that

$$f(\boldsymbol{\mathcal{S}}\lambda') \le f(\boldsymbol{\mathcal{D}}\lambda') \tag{3.15}$$

for some stochastic matrix $\boldsymbol{\mathcal{D}}$, which can be decomposed into a series of permutation matrices $\boldsymbol{\mathcal{P}}_\alpha$ via Birkhoff's theorem as defined in (3.11), i.e.,

$$f(\boldsymbol{\mathcal{D}}\lambda') = f\Big(\sum_\alpha \omega_\alpha \boldsymbol{\mathcal{P}}_\alpha \lambda'\Big) \qquad \text{with} \qquad \omega_\alpha \ge 0 \qquad \text{and} \qquad \sum_\alpha \omega_\alpha = 1. \tag{3.16}$$

The function $f$ is convex by definition and consequently we can make use of Jensen's inequality, such that

$$f\Big(\sum_\alpha \omega_\alpha \boldsymbol{\mathcal{P}}_\alpha \lambda'\Big) \le \sum_\alpha \omega_\alpha f(\boldsymbol{\mathcal{P}}_\alpha \lambda'). \tag{3.17}$$

Furthermore, $f$ is invariant under permutation of its arguments, i.e., $f(\lambda) = f(\boldsymbol{\mathcal{P}}\lambda)$, and we have

$$\sum_\alpha \omega_\alpha f(\boldsymbol{\mathcal{P}}_\alpha \lambda') = \sum_\alpha \omega_\alpha f(\lambda') = f(\lambda'). \tag{3.18}$$

Combining (3.14)–(3.18), we arrive at

$$f(\lambda) \le f(\lambda') = f\big(t\,\lambda_1 + (1-t)\,\lambda_2\big), \tag{3.19}$$

which is the desired result.[10] In the process, we have made use of all three property of $f$, i.e., its convexity, monotonicity, and permutation-invariance. □

Proposition 3.2 can be applied to the first and last three arguments of $g(\lambda, \boldsymbol{a})$ separately which then implies (3.5).

### 3.1.4. *Implications for Hill's inequality*

As discussed in (2.11), we can ensure Hill's inequality straightforwardly by requiring the strain-energy function $W(\mathbf{F}) = \widehat{W}(\log\lambda_1, \log\lambda_2, \log\lambda_3)$ to be strictly convex in the logarithmic strain $\log\mathbf{V}$. We reiterate that in case of incompressibility this strict convexity is only sufficient, not necessary, as discussed following (2.8).

As will be shown in just a moment, Ball's conditions for polyconvexity in Theorem 3.1 also (almost) ensure strict convexity in $\log\mathbf{V}$ and therefore Hill's inequality which implies a particularly economic parametrization of $W$ in terms of the principal stretches. To the authors' knowledge, this result has not yet been stated explicitly in the literature.

[10] The structure of the proof has much in common with the arguments in Ball (1976, p. 365), although the theory of majorization and Birkhoff's theorem are not mentioned there.

Notice that (3.2) and $\det \mathbf{F} = \lambda_1\lambda_2\lambda_3 = 1$ imply

$$W(\mathbf{F}) = g(\lambda_1, \lambda_2, \lambda_3, \lambda_2\lambda_3, \lambda_3\lambda_1, \lambda_1\lambda_2) = g\left(\lambda_1, \lambda_2, \lambda_3, \lambda_1^{-1}, \lambda_2^{-1}, \lambda_3^{-1}\right) \tag{3.20}$$

and further

$$\widehat{W}(\log \lambda_1, \log \lambda_2, \log \lambda_3) = g\left(e^{\log \lambda_1}, e^{\log \lambda_2}, e^{\log \lambda_3}, e^{-\log \lambda_1}, e^{-\log \lambda_2}, e^{-\log \lambda_3}\right). \tag{3.21}$$

As noted by Hill (1968, p. 238), for $\widehat{W}$ to be strictly convex in $\log \mathbf{V}$ it is necessary and sufficient for the strain-energy function to be strictly convex in the logarithmic principal stretches $\log \lambda_k$, cf. also Ball (1976, Thm. 5.1) and d'Agostino et al. (2025, App. A.4.1). Consequently, strict convexity of $g$ in $\log \lambda_k$ is sufficient for Hill's inequality in case of incompressibility which leads us to the following proposition:

**Proposition 3.3.** *Let $W$ satisfy the conditions for incompressible polyconvexity defined in Theorem 3.1. Additionally, let the monotonicity in either the first or last three arguments of $g$ be strict. Then $W$ satisfies Hill's inequality.*

**Proof.** By virtue of Theorem 3.1 and incompressibility, we have

$$\begin{aligned} W(\mathbf{F}) &= \widehat{W}(\log \lambda_1, \log \lambda_2, \log \lambda_3) \\ &= g\left(e^{\log \lambda_1}, e^{\log \lambda_2}, e^{\log \lambda_3}, e^{-\log \lambda_1}, e^{-\log \lambda_2}, e^{-\log \lambda_3}\right), \end{aligned} \tag{3.22}$$

where $g$ is convex and monotonically increasing in its six arguments, here strictly so in either its first or last three arguments. Since the exponential function is itself strictly convex, the strict convexity of $\widehat{W}$ in $\log \mathbf{V}$ follows immediately.

To make this more explicit, we introduce $\boldsymbol{h} = (\log \lambda_k)_{k=1}^3$ and define $\exp \boldsymbol{h} = (\exp h_k)_{k=1}^3$ such that

$$W(\mathbf{F}) = g\left(\exp \boldsymbol{h}, \exp(-\boldsymbol{h})\right). \tag{3.23}$$

The exponential function is strictly convex, hence

$$\exp\left(t\, h_1 + (1-t)h_2\right) < t \exp h_1 + (1-t)\exp h_2 \quad \forall h_1 \neq h_2 \quad \forall t \in [0,1]. \tag{3.24}$$

With $\boldsymbol{h} = t\,\boldsymbol{h}_1 + (1-t)\,\boldsymbol{h}_2$, we have

$$\begin{aligned} g\left(\exp \boldsymbol{h}, \exp(-\boldsymbol{h})\right) &= g\Big(\exp\left(t\,\boldsymbol{h}_1 + (1-t)\,\boldsymbol{h}_2\right), \exp\left(-t\,\boldsymbol{h}_1 - (1-t)\,\boldsymbol{h}_2\right)\Big) \\ &< g\left(t \exp \boldsymbol{h}_1 + (1-t)\exp \boldsymbol{h}_2, t\exp(-\boldsymbol{h}_1) + (1-t)\exp(-\boldsymbol{h}_2)\right) \\ &\leq t\, g\left(\exp \boldsymbol{h}_1, \exp(-\boldsymbol{h}_1)\right) + (1-t)\, g\left(\exp \boldsymbol{h}_2, \exp(-\boldsymbol{h}_2)\right), \end{aligned} \tag{3.25}$$

which proves the strict convexity of $\widehat{W}$ in $\log \mathbf{V}$ and in turn Hill's inequality given $W$. The stricter upper bound in the second line follows from the strict monotonicity of $g$ in either its first or last three arguments and (3.24), while the upper bound in line three is the result of the convexity of $g$. □

**Remark 3.4.** The current result can be seen as a generalization of the observation by Wollner et al. (2026a, Prop. 5.10), where a similar argument was used, albeit solely for the specific deformation mode of unconstrained uniaxial extension-compression.

**Remark 3.5.** The argument (3.25) works analogously if we require strictness in the convexity of $g$ instead of the monotonicity.

### 3.2. Parametrization based on invariants

Another widespread choice of parametrization comes in the form of 'strain' invariants of, e.g., the principal invariants of right Cauchy–Green tensor $\mathbf{C} = \mathbf{F}^{\mathrm{T}}\mathbf{F}$.[11] On a surface level, the parametrization in terms of invariants or principal stretches is simply a matter of preference. However, from the viewpoint of polyconvexity and Hill's inequality, the latter seems more powerful as many results for invariant-based strain-energy functions follow directly from Theorems 3.1 and Proposition 3.3 which are formulated using $\lambda_k$. This can be seen as follows:

**Corollary 3.6.** *Assuming incompressibility, let $W(\mathbf{F}) = \Psi(K_1, K_2)$, where the invariants $K_1$ and $K_2$ are Schatten $p$-norms in $\mathbf{F}$ and $\operatorname{Cof}\mathbf{F}$, respectively.*[12] *If $\Psi$ is convex and monotonically increasing, strictly so in either $K_1$ or $K_2$, then $W$ is polyconvex and satisfies Hill's inequality.*

[11] The term 'strain' invariants is a slight misnomer if the invariants are derived from a tensor such as $\mathbf{C}$ which does not vanish in the undeformed configuration. Technically, the principal stretches $\lambda_k$ are also tensor invariants and both parametrizations are in many ways interchangeable, albeit not always efficiently so, cf. Sawyers (1986) and Rivlin (2006). However, we employ this casual notation as we believe it to improve readability of the text without leaving much room for confusion in isotropic hyperelasticity.

[12] Physically, the invariants $K_1$ and $K_2$ can be understood as a generalized metric measuring the deformation of line and area elements, respectively, cf. Kearsley (1989). Furthermore, in case of incompressibility, we can observe an interesting symmetry between the expressions for $K_1$ in $\lambda_k$ and $K_2$ in $\lambda_k^{-1}$, cf. also Rivlin (2006, Eq. (36)).

**Proof.** By definition, the Schatten $p$-norm of $\mathbf{F}$ reads

$$\|\mathbf{F}\|_p = \left(\sum_{k=1}^{3} \lambda_k(\mathbf{F})^p\right)^{1/p} \quad \forall\, p \in [1, \infty), \tag{3.26}$$

i.e., the $p$-norm of vector of the principal stretches, cf. Horn and Johnson (1985, p. 441). It is straightforward to show that $\|\mathbf{F}\|_p$ is strictly increasing in $\lambda_k$ by differentiation. Similarly, convexity in $\lambda_k$ follows directly from the basic properties of vector norms. Observe that (3.26) is invariant under reordering of the principal stretches.

If we define

$$g(\boldsymbol{\lambda}, \boldsymbol{a}) = \Psi\big(\|\mathbf{F}\|_p, \|\mathbf{A}\|_q\big), \tag{3.27}$$

where $\Psi(K_1, K_2)$ is convex and monotonically increasing, strictly so in either $K_1$ or $K_2$, then $g$ is:

(i) convex,
(ii) monotonically increasing, strictly so in either $\boldsymbol{\lambda}$ or $\boldsymbol{a}$,
(iii) invariant under separate permutation of $\boldsymbol{\lambda}$ and $\boldsymbol{a}$.

Then it follows from Theorems 3.1 and Proposition 3.3 that $W$ is polyconvex and satisfies Hill's inequality. □

This novel result is interesting in so far as it generalizes a number of previous results. For $p = q = 1$, we have

$$K_1 = \|\mathbf{F}\|_1 = \lambda_1 + \lambda_2 + \lambda_3 \quad \text{and} \quad K_2 = \|\mathrm{Cof}\,\mathbf{F}\|_1 = \lambda_1\lambda_2 + \lambda_2\lambda_3 + \lambda_3\lambda_1 = \lambda_1^{-1} + \lambda_2^{-1} + \lambda_3^{-1} \tag{3.28}$$

which are the principal invariants of $\mathbf{V}$ which have been used by Steigmann (2003) in the context of polyconvexity, cf. also Wiedemann and Peter (2026, Sect. 5).

For $p = q = 2$, we recover

$$K_1 = \|\mathbf{F}\|_2 = \sqrt{\lambda_1^2 + \lambda_2^2 + \lambda_3^2} \quad \text{and} \quad K_2 = \|\mathrm{Cof}\,\mathbf{F}\|_2 = \sqrt{(\lambda_1\lambda_2)^2 + (\lambda_2\lambda_3)^2 + (\lambda_3\lambda_1)^2} = \sqrt{\lambda_1^{-2} + \lambda_2^{-2} + \lambda_3^{-2}}, \tag{3.29}$$

which have recently been employed in an investigation by Wollner et al. (2026a) concerning polyconvexity and TSTS-M.[13] The results on isotropic polyconvexity by Schröder and Neff (2003) and Hartmann and Neff (2003) primarily utilize the classical principal invariants of $\mathbf{C}$, i.e.,

$$I_1 = \mathrm{tr}\,\mathbf{C} = K_1^2 \quad \text{and} \quad I_2 = \mathrm{tr}\,\mathrm{Cof}\,\mathbf{C} = K_2^2, \tag{3.30}$$

and are implied by the Corollary 3.6. Consequently, a convex, monotonically increasing neural network, parametrized in $I_1$ and $I_2$, has by definition less approximation power compared to an analogous PANN parametrized in $K_1$ and $K_2$ since the square root operation cannot be approximated by the former.

Particularly noteworthy is the relation to the Ogden model, cf. Ogden (1972) and somewhat related Anssari-Benam et al. (2024). Focusing on just one Ogden term, we have

$$W(\mathbf{F}) = \frac{\mu}{p}(\lambda_1^p + \lambda_2^p + \lambda_3^p - 3). \tag{3.31}$$

Notice that

$$W(\mathbf{F}) = \frac{\mu}{p}(\lambda_1^p + \lambda_2^p + \lambda_3^p - 3) \quad \Longrightarrow \quad \Psi(K_1, K_2) = \begin{cases} \dfrac{\mu}{p} K_1^p - 3, & \text{if } p \geq 1, \\ \dfrac{\mu}{p} K_2^p - 3, & \text{if } p \leq -1, \end{cases} \tag{3.32}$$

since

$$K_1^p = \|\mathbf{F}\|_p^p = \lambda_1^p + \lambda_2^p + \lambda_3^p \quad \text{and} \quad K_2^p = \|\mathrm{Cof}\,\mathbf{F}\|_p^p = (\lambda_1\lambda_2)^p + (\lambda_2\lambda_3)^p + (\lambda_1\lambda_3)^p = \lambda_1^{-p} + \lambda_2^{-p} + \lambda_3^{-p}. \tag{3.33}$$

In both cases $\Psi$ is convex and strictly increasing in either $K_1$ or $K_2$ if $\mu\, p > 0$ and $|p| \geq 1$ such that the Ogden model is polyconvex and satisfies Hill's inequality. Obviously, this is in itself no new result: (i) the proof of polyconvexity can already be found in Ball (1976, Sect. 8) and (ii) the proof of Hill's inequality is given in Ogden (1972, Sect. 6). It is nonetheless instructive to see that these results are all implied by Corollary 3.6.

[13] Corollary 3.6 generalizes the result by Wollner et al. (2026a, Rem. 5.12), where $p = q = 2$. The proof here has fewer restrictions concerning the differentiability of $W$ and notably does not require the additional constraint defined in Wollner et al. (2026a, Eq. (5.58)). The particular set of invariants has already been used by Renardy (1985, Lem. 2.1) in a convexity argument. That the square root operation can, in specific cases, preserve convexity of a function has also been made use of in $J_2$-plasticity, cf. Simo and Hughes (1998, Eq. (2.5.1) and Prop. 2.6.1).

**Remark 3.7.** For $W(\mathbf{F}) = \widetilde{\Psi}(I_1, I_2)$, the Cauchy stress response follows as

$$\sigma = 2\frac{\partial\widetilde{\Psi}}{\partial I_1}\mathbf{b} - 2\frac{\partial\widetilde{\Psi}}{\partial I_2}\mathbf{b}^{-1} - \tilde{p}\,\mathbb{1}, \tag{3.34}$$

where $\tilde{p}$ is an arbitrary contribution to the hydrostatic pressure and $\mathbf{b} = \mathbf{F}\,\mathbf{F}^{\mathrm{T}}$ denotes the left Cauchy–Green tensor. Truesdell's empirical inequalities, in case of incompressible hyperelasticity, require that

$$\frac{\partial\widetilde{\Psi}}{\partial I_1} > 0 \quad \text{and} \quad \frac{\partial\widetilde{\Psi}}{\partial I_2} \geq 0 \quad \text{or} \quad \frac{\partial\widetilde{\Psi}}{\partial I_1} \geq 0 \quad \text{and} \quad \frac{\partial\widetilde{\Psi}}{\partial I_2} > 0 \tag{3.35}$$

cf. Truesdell and Noll (1965, Sect. 53). These conditions are identical to the monotonicity requirements in Corollary 3.6, since

$$\frac{\partial\widetilde{\Psi}}{\partial I_1} = \frac{1}{2K_1}\frac{\partial\Psi}{\partial K_1} \quad \text{and} \quad \frac{\partial\widetilde{\Psi}}{\partial I_2} = \frac{1}{2K_2}\frac{\partial\Psi}{\partial K_2}, \tag{3.36}$$

by virtue of (3.30). However, the empirical inequalities make no statement about the second derivatives, i.e., the convexity of $\widetilde{\Psi}$. For example $\widetilde{\Psi}(I_1, I_2) = I_1^{1/4} - 3^{1/4}$ is strictly monotonically increasing, but not convex and has a non-monotonic true-shear-stress response in simple shear with

$$\sigma_{12}(\gamma) = \frac{\gamma}{2(3+\gamma^2)^{3/4}}, \tag{3.37}$$

which is indicative of a lack of rank-one convexity, cf. Wollner et al. (2026a, Sect. 5.3). Consequently, the empirical inequalities by themselves should be considered too weak as a constitutive requirement, see the implications in Fig. 1; on a related note, cf. also Thiel et al. (2019).

### *3.3. An open question: Does polyconvexity imply TSTS-M in case of incompressibility?*

Having established a number of results relating polyconvexity and TSTS-M in case of incompressibility, it is sensible to discuss some very general implications. To reiterate, *any implication discussed in this particular section is done within the constraint to incompressibility in mind.*

From Proposition 3.3, we have[14]

$$\text{Ball's conditions for polyconvexity} \quad \Longrightarrow \quad \text{convexity of } \widehat{W} \text{ in } \log\mathbf{V} \quad \Longrightarrow \quad \text{TSTS-M}. \tag{3.38}$$

The converse of the first implication is not true which can straightforwardly be seen by way of an implicit example. The original proof of Hill's inequality by Ogden (1972, Sect. 6) for the Ogden model in (3.31) does not place any limits on the exponent $p$. Hence,

$$W(\mathbf{F}) = \sqrt{\lambda_1} + \sqrt{\lambda_2} + \sqrt{\lambda_3} - 3 \tag{3.39}$$

with $p = \frac{1}{2}$ satisfies Hill's inequality. Since here $|p| < 1$, we cannot make use of Corollary 3.6 to also establish polyconvexity. In fact, $W$ is not even rank-one convex. To see this, notice that the true-shear-stress response in simple shear, resulting from (3.39), reads

$$\sigma_{12}(\gamma) = \frac{\left(\gamma + \sqrt{\gamma^2+4}\right)^{1/2} - \left(-\gamma + \sqrt{\gamma^2+4}\right)^{1/2}}{2^{3/2}\sqrt{\gamma^2+4}}, \tag{3.40}$$

which is non-monotonic, cf. Zee and Sternberg (1983, Fig. 4). Consequently, $W$ is also not polyconvex.

The converse of the second implication in (3.38) also does not hold which we already discussed following (2.12). We only require convexity of $\widehat{W}^{\mathrm{inc}}_{\mathrm{red}}$ in (2.13) to satisfy the weak form of Hill's inequality. This in turn asks the question, if there exists a polyconvex strain-energy function that does not satisfy Ball's conditions, but nonetheless entails TSTS-M. One possible avenue to an answer are given by the necessary and sufficient condition for polyconvexity recently summarized by Wiedemann and Peter (2026, Thm. 2.1) and Geuken et al. (2026, Cor. 2). Take for example the incompressible strain-energy function

$$W(\mathbf{F}) = (\lambda_1 - \lambda_2\lambda_3)^4 + (\lambda_2 - \lambda_1\lambda_3)^4 + (\lambda_3 - \lambda_1\lambda_2)^4 \tag{3.41}$$

with

$$g(\lambda_1, \lambda_2, \lambda_3, a_1, a_2, a_3) = (\lambda_1 - a_1)^4 + (\lambda_2 - a_2)^4 + (\lambda_3 - a_3)^4, \tag{3.42}$$

inspired by Mielke (2005, Eq. (20)). The function $g$ is $\Pi(3)$-invariant[15] and convex over $\mathbb{R}^6$, hence $W$ is polyconvex, even though Ball's conditions do not hold for the specific definition of $g$ in (3.42) since

$$\frac{\partial g}{\partial \lambda_1} = 4(\lambda_1 - a_1)^3 \tag{3.43}$$

[14] Here, we deliberately neglect the nuance associated with the fact that TSTS-M is a strict inequality, while polyconvexity is only weak.

[15] $\Pi(3)$-invariance means that the function $g(\lambda, \boldsymbol{a})$ is invariant under shared permutation of both $\lambda_k$ and $a_k$, e.g., $g(\lambda_1, \lambda_2, \lambda_3, a_1, a_2, a_3) = g(\lambda_3, \lambda_1, \lambda_2, a_3, a_1, a_2)$, as well as shared reflection of two entries of $\lambda$ and $\boldsymbol{a}$, e.g., $g(\lambda_1, \lambda_2, \lambda_3, a_1, a_2, a_3) = g(\lambda_1, -\lambda_2, -\lambda_3, a_1, -a_2, -a_3)$, cf. Wiedemann and Peter (2026, Sect. 2.1).

is not always positive for all $\lambda_1, a_1 \in \mathbb{R}_+$. By expanding each term and using $\lambda_1\lambda_2\lambda_3 = 1$, we may rewrite (3.41) in the form

$$\begin{aligned} W(\mathbf{F}) &= (\lambda_1 - \lambda_2\lambda_3)^4 + (\lambda_2 - \lambda_1\lambda_3)^4 + (\lambda_3 - \lambda_1\lambda_2)^4 \\ &= \lambda_1^2(\lambda_1^2-4) + \lambda_2^2(\lambda_2^2-4) + \lambda_3^2(\lambda_3^2-4) + \lambda_2^2\lambda_3^2(\lambda_2^2\lambda_3^2-4) + \lambda_1^2\lambda_3^2(\lambda_1^2\lambda_3^2-4) + \lambda_1^2\lambda_2^2(\lambda_1^2\lambda_2^2-4) + 18, \end{aligned} \tag{3.44}$$

which allows the new assignment[16]

$$\tilde{g}(\lambda_1,\lambda_2,\lambda_3,a_1,a_2,a_3) = \lambda_1^2(\lambda_1^2-4) + \lambda_2^2(\lambda_2^2-4) + \lambda_3^2(\lambda_3^2-4) + a_1^2(a_1^2-4) + a_2^2(a_2^2-4) + a_3^2(a_3^2-4) + 18. \tag{3.45}$$

However, this choice does not provide a path to polyconvexity by Ball's conditions either.

Taking a look at the parametrization in terms of the Hencky strain, we have

$$\widehat{W}(\log\lambda_1,\log\lambda_2,\log\lambda_3) = \left(e^{\log\lambda_1} - e^{\log\lambda_2+\log\lambda_3}\right)^4 + \left(e^{\log\lambda_2} - e^{\log\lambda_1+\log\lambda_3}\right)^4 + \left(e^{\log\lambda_3} - e^{\log\lambda_1+\log\lambda_2}\right)^4, \tag{3.46}$$

which is not convex in $\log\lambda_k$ since

$$\left.\frac{\partial^2\widehat{W}}{\partial(\log\lambda_1)^2}\right|_{\substack{\lambda_1=\lambda\\ \lambda_2=\lambda^{-1}\\ \lambda_3=1}} = \frac{8(\lambda^2-1)^2(4\lambda^2-1)}{\lambda^2} < 0 \qquad \forall\lambda < \frac{1}{2}. \tag{3.47}$$

However, with (2.13), we have

$$\begin{aligned} \widehat{W}^{\text{inc}}_{\text{red}}(\log\lambda_1,\log\lambda_2) &= \widehat{W}\left(\log\lambda_1,\log\lambda_2,-\log\lambda_1-\log\lambda_2\right) \\ &= \left(e^{\log\lambda_1} - e^{-\log\lambda_1}\right)^4 + \left(e^{\log\lambda_2} - e^{-\log\lambda_2}\right)^4 + \left(e^{-\log\lambda_1-\log\lambda_2} - e^{\log\lambda_1+\log\lambda_2}\right)^4 \\ &= 16\sinh^4(\log\lambda_1) + 16\sinh^4(\log\lambda_2) + 16\sinh^4(\log\lambda_1+\log\lambda_2), \end{aligned} \tag{3.48}$$

where each term is strictly convex in $\log\lambda_1$ and $\log\lambda_2$. Hence, $W$ in (3.41) is both polyconvex and satisfies the weak form of Hill's inequality and, in turn, TSTS-M, although Ball's conditions do not apply. Whether polyconvexity implies TSTS-M in case of incompressibility in general is another matter, though.[17] Further discussion lies outside the scope of this contribution and will be addressed elsewhere.

**Remark 3.8.** We want to reiterate that, in case of compressibility, polyconvexity does not imply TSTS-M, cf. Leblond (1992, Sect. 4b), even if Ball's sufficient conditions are employed, cf. Wollner et al. (2026a, Prop. 5.7). Thus, an extension of our results to the general compressible case is not possible. Moreover, TSTS-M is first and foremost a condition restricted to isotropic material behavior. Although a generalization to anisotropy is technically possible, the condition loses much of its mathematical simplicity. More importantly, it is considerably less intuitive from a physical perspective to require some sort of multiaxial monotonicity condition in case of anisotropy. This is not to say that investigations along such lines could not be of interest.

## 4. PANN constitutive modeling

We now demonstrate how the above introduced framework can be transferred to physics-augmented neural network (PANN) constitutive models, for which we employ multiple established PANN models from the literature. For the PANN model formulation, we explicitly take polyconvexity into account, while the TSTS-M condition then follows from above given theoretical discussions. The latter part does not apply to the PANN based on signed singular values. We introduce the constitutive model based on neural networks in Section 4.1, followed by stress predictions for simple load paths in Section 4.3. The intermediate Section 4.2 briefly discusses the conceptual differences between classical and neural-network-based approaches.

### 4.1. Neural networks as constitutive model equations

In hyperelastic PANN constitutive modeling, feedforward neural networks (FFNNs) are used to represent the strain-energy function. While a variety of different FFNN architectures could be employed for this, in practical applications, a rather simple architecture has proven to be sufficiently flexible across various scenarios reading

$$f:\mathbb{R}^m\to\mathbb{R}, \qquad \boldsymbol{x}\mapsto f(\boldsymbol{x}) = \sum_{k=1}^{n}\sum_{l=1}^{m} w_k\,\mathrm{SP}(W_{kl}x_l + b_k) = \left\langle \boldsymbol{w}, \mathrm{SP}(\mathcal{W}\boldsymbol{x}+\boldsymbol{b})\right\rangle, \tag{4.1}$$

cf. Klein et al. (2024) and Kalina et al. (2025). In the language of machine learning, relation (4.1) constitutes a FFNN with a single hidden layer using the softplus activation function $\mathrm{SP}(x) = \log(1+e^x)$. The weight matrices $\mathcal{W}\in\mathbb{R}^{n\times m}, \boldsymbol{w}\in\mathbb{R}^n$ and the bias vector $\boldsymbol{b}\in\mathbb{R}^n$ form the collection of parameters $\boldsymbol{\theta}$ that are optimized to fit the neural network to a given dataset. Here, $n$ is referred to as the amount of nodes in the FFNN. As $n$ increases, both the number of trainable parameters and the flexibility of the model grow. In the following, we consider the special FFNN architectures, originally proposed by Amos et al. (2017): (i) input-convex neural networks (ICNNs) which are convex in their arguments and (ii) convex-monotonic neural networks (CMNNs) which are convex and monotonically non-decreasing in their arguments. For the special FFNN in (4.1), this can be included with the following conditions:

[16] We would like to thank Gian-Luca Geuken for bringing this alternative assignment to our attention.

[17] In the incompressible two-dimensional case, polyconvexity does in fact imply TSTS-M, cf. Ghiba et al. (2027).

The softplus function is convex and monotonic. Thus, if the weights in $\boldsymbol{w}$ are non-negative, then $f$ becomes an ICNN. Moreover, if the weights in $\boldsymbol{w}$ and $\mathcal{W}$ are non-negative, then $f$ becomes a CMNN.[18]

Based on (4.1), polyconvex strain-energy functions can be constructed, which was originally proposed by Klein et al. (2022) for invariant-based modeling. For principal stretch-based modeling, multiple approaches have been proposed that deviate in how they incorporate permutation invariance, which can be achieved either with a weight sharing strategy, as proposed by Vijayakumaran et al. (2024), or with a group symmetrization strategy, as proposed by Geuken et al. (2025).[19] In the latter work, also a signed-singular-value-based PANN model has been proposed, in which the corresponding invariance conditions are likewise incorporated via a group symmetrization approach. In this work, we consider the following polyconvex architectures:[20]

- **PANN based on invariants**: Let $\boldsymbol{I}^{\square}$ be a tuple of invariants, where $\square$ is a placeholder denoting a specific choice of Schatten $p$-norms. Consequently, using $\boldsymbol{I}^{\square}$ as inputs for the CMNN

$$\Psi_{\square}^{\mathrm{NN}} = \left\langle \boldsymbol{w}, \mathrm{SP}(\mathcal{W}\boldsymbol{I}^{\square} + \boldsymbol{b})\right\rangle \qquad \text{with} \qquad w_k, W_{kl} \geq 0, \tag{4.2}$$

  yields a polyconvex PANN by virtue of Corollary 3.6. We consider two different sets of invariants, resulting in two distinct neural networks: (i) one based on the principal invariants of the right Cauchy–Green tensor with $\boldsymbol{I}^{\mathrm{st}} = (\mathrm{tr}\,\mathbf{C}, \mathrm{tr}\,\mathrm{Cof}\,\mathbf{C})$ and (ii) another based on the respective square root invariants $\boldsymbol{I}^{\mathrm{sr}} = (\sqrt{\mathrm{tr}\,\mathbf{C}}, \sqrt{\mathrm{tr}\,\mathrm{Cof}\,\mathbf{C}})$. We refer to the first and second model as PANN–$\boldsymbol{I}$ and PANN–$\sqrt{\boldsymbol{I}}$, respectively.[21] Both architectures additionally satisfy TSTS-M *a priori* due to Corollary 3.6. This specific PANN architecture was originally proposed by Klein et al. (2022).
- **PANN based on principal stretches**: Let $\boldsymbol{x}_\lambda = (\lambda_1, \lambda_2, \lambda_3, \lambda_2\lambda_3, \lambda_3\lambda_1, \lambda_1\lambda_2)$ be a tuple containing the principal stretches $\lambda_k$. When $\boldsymbol{x}_\lambda$ is used as input of a CMNN that is permutation-invariant in $\lambda_k$, a polyconvex strain-energy function is obtained by virtue of Theorem 3.1. Hence,

$$\Psi_{\lambda}^{\mathrm{NN}} = \frac{1}{6}\sum_{\mathcal{P}} \left\langle \boldsymbol{w}, \mathrm{SP}(\mathcal{W}\mathcal{P}\boldsymbol{x}_\lambda + \boldsymbol{b})\right\rangle \qquad \text{with} \qquad w_k, W_{kl} \geq 0, \tag{4.3}$$

  where we sum over the six possible permutations of $\lambda_k$, encapsulated by the permutation matrix $\mathcal{P}$, ensures the permutation invariance of the strain-energy function, cf. Fernández et al. (2022, Eq. (11)). Since $\mathcal{P}\boldsymbol{x}_\lambda$ is a linear operation, it preserves the convexity. We refer to this model as PANN–$\lambda$. With Proposition 3.3, this particular PANN also satisfies TSTS-M *a priori*. The architecture was originally proposed by Geuken et al. (2026, Sect. 4.3).
- **PANN based on signed singular values**: Let $\boldsymbol{x}_\nu = (\nu_1, \nu_2, \nu_3, \nu_2\nu_3, \nu_3\nu_1, \nu_1\nu_2)$ be a tuple containing the signed singular values $\nu_k$. When $\boldsymbol{x}_\nu$ is used as input of a convex function that is $\Pi(3)$-invariant, a polyconvex strain-energy function is obtained by virtue of Geuken et al. (2026, Cor. 2). In an ICNN architecture, originally proposed by Geuken et al. (2025), we have

$$\Psi_{\nu}^{\mathrm{NN}} = \frac{1}{24}\sum_{\mathcal{P}_{\mathrm{sgn}}} \left\langle \boldsymbol{w}, \mathrm{SP}(\mathcal{W}\mathcal{P}_{\mathrm{sgn}}\boldsymbol{x}_\nu + \boldsymbol{b})\right\rangle \qquad \text{with} \qquad w_k \geq 0, \tag{4.4}$$

  where $\mathcal{P}_{\mathrm{sgn}}$ are the 24 signed permutation matrices belonging to $\Pi(3)$, cf. Geuken et al. (2025, Table 1). Again, $\mathcal{P}_{\mathrm{sgn}}\boldsymbol{x}_\nu$ is a linear operation and thus preserves convexity. We refer to this model as PANN–$\nu$. In contrast to the other previous architectures, it is not certain that TSTS-M is guaranteed *a priori*, as discussed in Section 3.3.

The investigated PANN models are similar in that they satisfy (almost) all relevant constitutive constraints *a priori*. This ensures that the predicted material response remains mechanically reasonable even for large extrapolation away from the calibration data, at least in the sense of the considered constitutive constraints. In contrast, PANN models that do not satisfy all relevant constitutive constraints *a priori* often yield unphysical results for extrapolation, e.g., cf. Klein et al. (2022, 2026a). In addition, the inclusion of constitutive constraints in the model architecture can reduce the amount of data required for model calibration, cf. Linden et al. (2023), Kalina et al. (2024).

**Remark 4.1.** When the PANN–$\boldsymbol{I}/\sqrt{\boldsymbol{I}}$ models are evaluated for one specific deformation gradient, a single evaluation of the neural network is required. This is in contrast to the multiple evaluations required for the permutation invariances of the PANN–$\lambda/\nu$ models, which entail a higher computational cost.

### 4.2. On 'classical' and 'neural-network-based' constitutive models

The distinction between 'classical' and 'neural-network-based' constitutive models is somewhat artificial: FFNNs are merely mathematical functions that we employ to represent constitutive equations. As with classical approaches, we have closed-form mathematical expressions, e.g., (4.1), and can calculate all required derivatives analytically, cf. (Franke et al., 2023, Sect. 3.3). Moreover, both approaches are subject to the same mechanical considerations on constitutive constraints. Simply put, instead of a polynomial form linked to names such as Ogden, Mooney, and Rivlin, PANN models simply employ ansatz functions inspired by the machine learning community.

[18] For a general introduction to FFNNs, the reader is referred to Herrmann et al. (2025, Chap. 3). A discussion of multi-layered FFNNs for hyperelastic constitutive modeling can be found in Linden et al. (2023, App. A). Multi-layered FFNNs can also be formulated as ICNNs or CMNNs, cf. Klein et al. (2022, App. A) for explicit proofs.

[19] The concept of group symmetrization has also been employed in anisotropic PANN constitutive modeling, cf. Fernández et al. (2021), and is closely related to the *Reynolds operator* employed in invariant theory, cf. Derksen and Kemper (2015, Sect. 3.1.2).

[20] Note that there are scenarios, in which the considered material behavior practically precludes the inclusion of *all* constitutive constraints in the model formulation; this can be the case particularly for polyconvexity, cf. Klein et al. (2026b).

[21] While the invariants $\boldsymbol{I}^{\mathrm{sr}}$ have recently been employed by Abdolazizi et al. (2025, Sect. 2.1) for purposes of PANN constitutive modeling, the invariants $\boldsymbol{I}^{\mathrm{st}}$ have so far been a more common choice, e.g., cf. Taç et al. (2023) or Kalina et al. (2025).

The key distinction from classical models lies in the immediate increase in flexibility for the approach based on neural networks through increased size of the weight and bias matrices, i.e., by using more nodes in the hidden layer. Furthermore, multiple hidden layers can be employed. The neural network ansatz in (4.1) has demonstrated high stability during calibration, even when involving a large number of parameters, cf. Gärtner et al. (2021), Linka et al. (2021), and Kalina et al. (2025). In this matter, neural-network-based constitutive models are known to be very robust against phenomena such as overfitting which are known to occur for other neural network applications, cf. Herrmann et al. (2025). This is due to the fact that the incorporation of physical principles provides the model with a pronounced mathematical structure, cf. Linden et al. (2023) and Klein et al. (2026a). Furthermore, the neural network ansatz in (4.1) enables strong interdependencies among its input. This is a particular distinction from conventional polyconvex models. There, the potential is usually additively decomposed, e.g., in invariants, cf. Ebbing (2010), because the multiplication of polyconvex terms generally does not preserve polyconvexity, cf. Schröder and Neff (2003, Lem. C.1). In contrast, neural networks enable a pronounced coupling between the inputs, while simultaneously maintaining polyconvexity.

This improved flexibility comes at a cost: Compared to many classical models, PANNs are less interpretable, cf. Klein et al. (2022, Sect. 6), and can be computationally more expensive, cf. Franke et al. (2023, Sect. 7.2.1). Moreover, with some exceptions such as the signed singular value-based PANN model proposed by Geuken et al. (2025), most PANN models do *not* have universal approximation properties and cannot represent every material behavior. In many cases, the predictive capability of a constitutive model is limited by its general modeling framework, which can only in parts be compensated by very flexible ansatz functions. Examples include sufficient but not necessary monotonicity and convexity constraints in polyconvex constitutive modeling, cf. Klein et al. (2026b), or inelastic constitutive models based on dissipation potentials, which cannot represent every inelastic material behavior, cf. Masi (2026, Sect. 2.2.1).

To conclude, classical and neural-network-based constitutive models have more in common than what separates them: both employ closed-form constitutive equations, both are subject to the same constitutive constraints, and both have benefits and limitations. They mainly deviate in the choice of ansatz function made for the constitutive model equations.

### 4.3. *Stress prediction for simple load paths*

For actual stress computation in incompressible hyperelasticity, we require the overall constitutive model equations

$$\widetilde{W}:\ \mathrm{SL}(3)\times\mathbb{R}\to\mathbb{R},\qquad (\mathbf{F},\tilde{p})\mapsto W(\mathbf{F})-\tilde{p}\,(J-1),\qquad \text{and}\qquad \mathbf{S}_1=\mathrm{D}_{\mathbf{F}}\widetilde{W}(\mathbf{F},\tilde{p}), \tag{4.5}$$

that do not only depend on the strain-energy function $W$, but also on a Lagrange multiplier term $\tilde{p}\,(J-1)$ ensuring the incompressibility constraint. The tensor $\mathbf{S}_1$ denotes the first Piola–Kirchhoff stress. While $W$ is given by the specific choice of constitutive model, the Lagrange multiplier $\tilde{p}$ follows from the solution of a specific boundary-value problem. In this work, we evaluate the constitutive models for uniaxial tension (UX), equibiaxial tension (BX), and pure shear (PS). Assuming incompressibility, the deformation gradients for these load scenarios are given by

$$\mathbf{F}_{\mathrm{UX}}=\mathrm{diag}(\lambda,\lambda^{-1/2},\lambda^{-1/2}),\qquad \mathbf{F}_{\mathrm{BX}}=\mathrm{diag}(\lambda,\lambda,\lambda^{-2}),\qquad \text{and}\qquad \mathbf{F}_{\mathrm{PS}}=\mathrm{diag}(\lambda,1,\lambda^{-1}). \tag{4.6}$$

The corresponding stress tensor $\mathbf{S}_1$ is similarly diagonal, where the normal component along $\boldsymbol{e}_3$ and – in case of UX – along $\boldsymbol{e}_2$ vanishes. This allows us to explicitly calculate the Lagrange multiplier $\tilde{p}$ which provides the normal stress component of $\mathbf{S}_1$ along $\boldsymbol{e}_1$, denoted by $S_1$, with

$$S_1=\left(\frac{\partial W}{\partial\lambda_1}-\frac{\lambda_3}{\lambda_1}\frac{\partial W}{\partial\lambda_3}\right)\Big|_{\mathbf{F}_{\mathrm{UX/BX/PS}}}, \tag{4.7}$$

depending on the deformation mode, cf. Baaser (2026a, p. 7). Based on this, we will calibrate the PANN constitutive models in the next section.

## 5. Comparison of different PANN formulations applied to experimental data

We now apply the different PANN constitutive models introduced in the previous section to experimental data of soft rubber-like materials. We discuss the considered datasets and model calibration strategies in Section 5.1. This is followed by investigations on the model performance in inter- and extrapolation in Sections 5.2 and 5.3, respectively. In Section 5.4, we briefly revisit Truesdell's Hauptproblem in light of our numerical results.

### 5.1. *Considered datasets and model calibration*

We consider experimental data for three soft rubber-like materials: a vulcanized rubber by Treloar (1944), an EPDM polymer by Plagge and Klüppel (2017), and a 3D digital light processing material (DLP) by Zhang et al. (2024).[22] The first two materials are experimentally characterized by uniaxial tensile tests, equibiaxial tensile tests, and pure shear tests, while for the DLP material, only uniaxial tensile tests were carried out. For all materials, this results in datasets of the form

$$\mathcal{D} = \left\{ ({}^1\lambda,\, {}^1S_1), \dots, ({}^m\lambda,\, {}^mS_1) \right\}, \tag{5.1}$$

consisting of $m$ stretch-stress-parameter tuples, where $\lambda$ and $S_1$ denote the stretch and first Piola–Kirchhoff stress along the testing direction $\boldsymbol{e}_1$. To calibrate the parameters $\boldsymbol{\theta}$ of the neural network, the loss function, given as the mean squared error (MSE) with

$$\mathfrak{L}(\boldsymbol{\theta}) = \frac{1}{m_{\text{tot}}} \sum_{k=1}^{m_{\text{tot}}} \|{}^kS_1 - {}^kS_1^{\text{model}}({}^k\lambda; \boldsymbol{\theta})\|^2, \tag{5.2}$$

is minimized.[23] Here, $m_{\text{tot}}$ denotes the number of data points used for calibration and $S_1^{\text{model}}$ is evaluated according to (4.7). Note that we employ the first Piola–Kirchhoff stress for the evaluation of the MSE, but later on visualize the Cauchy stress response. The former stress measure has been shown to provide good results when calibrating PANN models to such experimental data, cf. Dammaß et al. (2025) and Klein et al. (2026a), while the latter is relevant for the TSTS-M condition. For each material, we employ all data points for calibration, meaning 56 data points for Treloar's data, 271 data points for the EPDM material, and 14 data points for the DLP material. The models are implemented and calibrated using Klax.[24] The parameter optimization is performed with the Adam optimizer, employing the full calibration dataset, a batch size of 32, without loss weighting, with a learning rate of 0.002, and for 150,000 calibration steps. In each investigation, the models are calibrated five times. By that, we account for the random initialization of the neural network parameters and stochastic effects in the optimization process. As discussed in the previous section, we employ the FFNN (4.1) to represent the strain-energy function for all PANN models with $n = 16$ nodes in the single hidden layer throughout. This results in 64 trainable parameters for the PANN–$\boldsymbol{I}/\sqrt{\boldsymbol{I}}$ models and 128 trainable parameters for the PANN–$\lambda/\nu$ models. In preliminary hyperparameter studies, this has shown to provide sufficient flexibility while maintaining a moderate number of trainable parameters. Note that once a neural network is sufficiently large for the targeted application, further increasing its size yields no notable advantage and only adds to the number of parameters, e.g., cf. Klein et al. (2026a, Sect. 4.2).

[22] In Zhang et al. (2024), multiple types of DLP are investigated which are synthesized under different manufacturing conditions. Here, we employ the DLP-50 dataset.

**Table 1**
Interpolation capabilities of the different polyconvex PANN models. The minimum mean squared error is reported logarithmically in $\log_{10}$ MPa as defined in (5.2).

| | PANN–$\boldsymbol{I}$ | PANN–$\sqrt{\boldsymbol{I}}$ | PANN–$\lambda$ | PANN–$\nu$ |
|---|---|---|---|---|
| Treloar | −2.42 | −2.79 | −2.80 | −2.77 |
| EPDM | −1.05 | −1.04 | −1.29 | −1.93 |
| DLP | 0.29 | 0.25 | −2.11 | −2.61 |

### 5.2. *Model performance in interpolation*

We begin by investigating the interpolation capabilities of the different PANN constitutive models.[25] The corresponding MSEs are provided in Table 1. For Treloar's data, the PANN–$\sqrt{\boldsymbol{I}}/\lambda/\nu$ models exhibit a very similar performance, whereas the PANN–$\boldsymbol{I}$ model yields a slightly higher MSE. For the EPDM dataset, the PANN–$\nu$ model clearly outperforms the other models, the PANN–$\lambda$ model achieves intermediate accuracy, and the PANN–$\boldsymbol{I}/\sqrt{\boldsymbol{I}}$ models show the largest MSEs. Similar trends are observed for the DLP material, however, the performance gap is more pronounced, with the invariant-based models performing significantly worse. To get an intuition about the quality of the stress response related to these MSE values, we exemplarily provide stress-stretch predictions for the PANN–$\boldsymbol{I}/\nu$ models in Fig. 2, which represent the lower and upper performance bounds among the considered PANN models. Although (partly significant) differences in the MSE values are observed for Treloar's and the EPDM data, their practical relevance appears limited. This is because in the stress-stretch curves, only minor differences between the model predictions are visible, and for these materials, all investigated models provide sufficiently accurate predictions. In contrast, for the DLP material, pronounced differences in the quality of stress prediction are observed: the PANN–$\boldsymbol{I}$ model shows clear deviations from the data, whereas the PANN–$\nu$ model shows an excellent performance. The reduced performance of the PANN–$\boldsymbol{I}$ model for the DLP material is not caused by a lack of model parameters in the neural network, but is due to the restrictions set by this choice of polyconvex parametrization. Overall, this investigation demonstrates that even when different PANN approaches are similar in that they satisfy the same constitutive conditions, they might differ in their flexibility which can influence their ability to represent some material behaviors.[26]

[23] The specific choice of a loss function can be argued on probabilistic ground, but is generally taken *ad hoc*, cf. Wollner et al. (2026b). Consequently, it should itself be seen as a modeling decision.

[24] Version 0.1, available under https://drenderer.github.io/klax/.

[25] Since we want to investigate the flexibility of the different PANN model approaches in terms of their interpolation capabilities, we employed the full datasets for calibration. For corresponding PANN model calibrations that do not rely on the full experimental dataset, the reader is referred to Taç et al. (2023), Dammaß et al. (2025), Abdolazizi et al. (2025), and Klein et al. (2026a).

[26] Similar observations are reported in Klein et al. (2022) for invariant- and coordinate-based models and in Taç et al. (2023) for models based on convex neural network architectures and on neural ordinary differential equations.

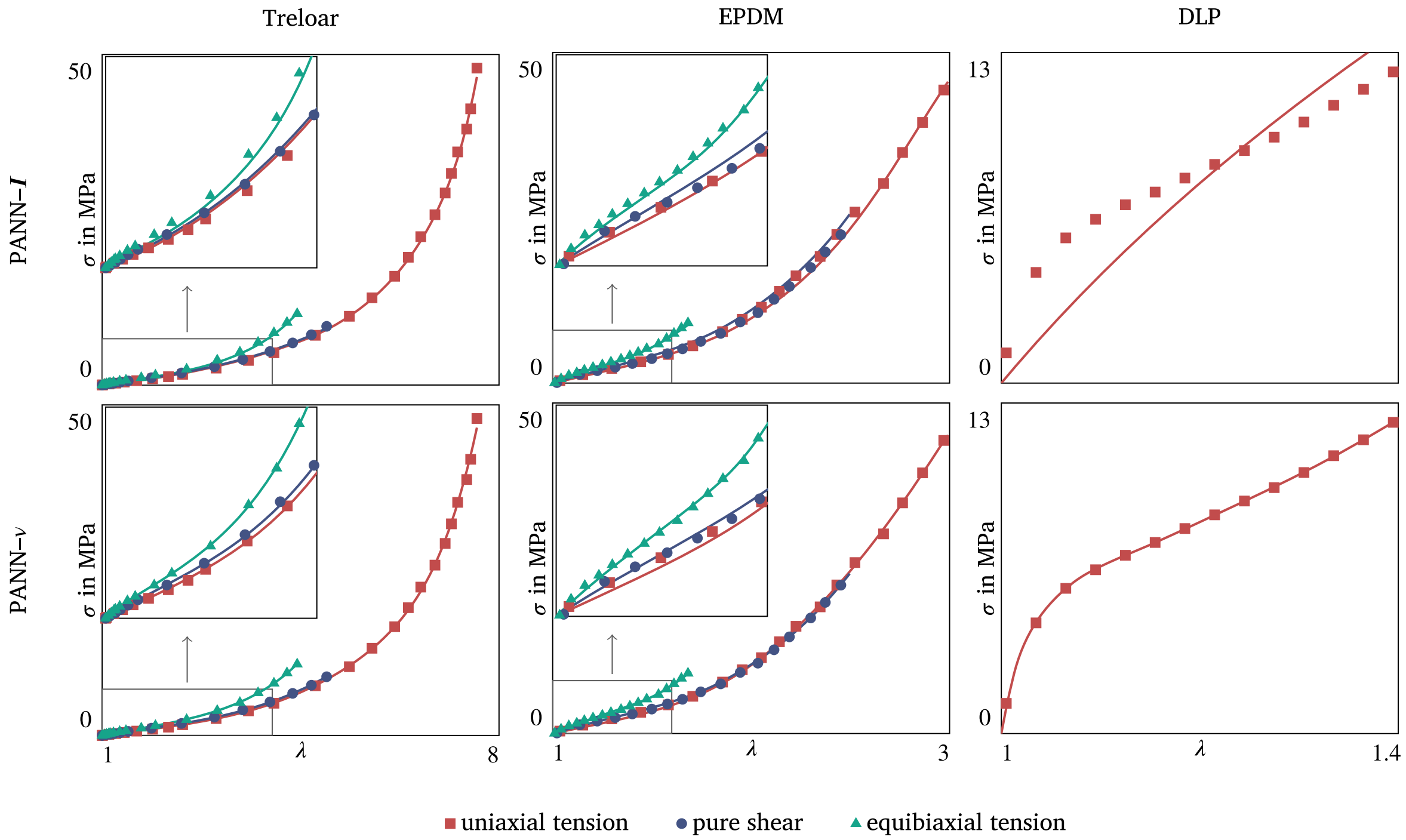

**Fig. 2.** Interpolation capabilities of the PANN–$\boldsymbol{I}$ and PANN–$\nu$. All deformation modes are used for calibration. Points label experimental data and solid lines indicate model predictions.

**Remark 5.1.** These results must further be interpreted in the context of the considered data and the underlying modeling assumptions. Experimental data are always interpretations in the light of an assumed material behavior and themselves inherently prone to measurement inaccuracies. Moreover, the assumption that the experiments correspond to idealized deformation states, e.g., equibiaxial tension, is only approximately satisfied. In addition, although we employ hyperelastic constitutive models, rubber-like materials are known to exhibit inelastic and rate-dependent effects. For instance, the EPDM material shows pronounced stress softening, cf. Plagge and Klüppel (2017). Furthermore, for Treloar's data and the DLP material, experiments were conducted at a single strain rate and without unloading, cf. Treloar (1944) and Zhang et al. (2024), such that potential inelastic and rate-dependent effects may be present but cannot be identified from the available data. These effects cannot be captured within the hyperelastic modeling framework considered here, which constitutes a clear limitation of our approach. Overall, both the data and the resulting model performance must therefore be interpreted with caution.

### 5.3. *Model performance in extrapolation*

We now investigate the extrapolation behavior of the different PANN constitutive models calibrated to Treloar's data.[27]

**Remark 5.2.** When investigating extrapolation away from the calibration data, the difference between multiple calibrated instances of the same architecture becomes more pronounced than in the interpolation regime. This is because the material parameter optimization problem is highly non-convex and the optimizer that we employ is stochastic, resulting in a different set of parameters each time a new model is calibrated. In our investigation, however, differently calibrated instances of the same model showed a very similar qualitative and quantitative performance in extrapolation, which we exemplarily show in Fig. 4. In the remaining figures, we only show the instance with the smallest associated MSE. Similar observations are reported in Klein et al. (2022, Sect. 4.3) and Klein et al. (2026a, Sect. 4.3). In the latter reference, it is also shown that without polyconvexity in the model formulation, the deviations between multiple calibrated instances of the same architecture become more pronounced.

In Fig. 3, we visualize the extrapolation behavior for the different PANN models as well as for two representative, conventional models. As expected, the Cauchy stress is monotonically increasing for all PANN models, even for large extrapolation away from the calibration data. While the different PANN models display a very similar performance in the interpolation regime, they show pronounced deviations both in the qualitative and quantitative stress response in the extrapolation regime. A similar behavior can be observed for the conventional constitutive models.

[27] We conducted similar investigations for PANN models calibrated to the EPDM and DLP datasets. As the results were similar and did not provide additional insights, they are omitted here for brevity. Large extrapolation of the constitutive model becomes particularly relevant when the models are employed in finite element analyses, where localized deformations often require evaluation of the constitutive model for very large strains. For the application of PANN constitutive models in finite element analysis, the reader is referred to Kalina et al. (2022), Franke et al. (2023), Klein et al. (2024), and Alheit et al. (2026).

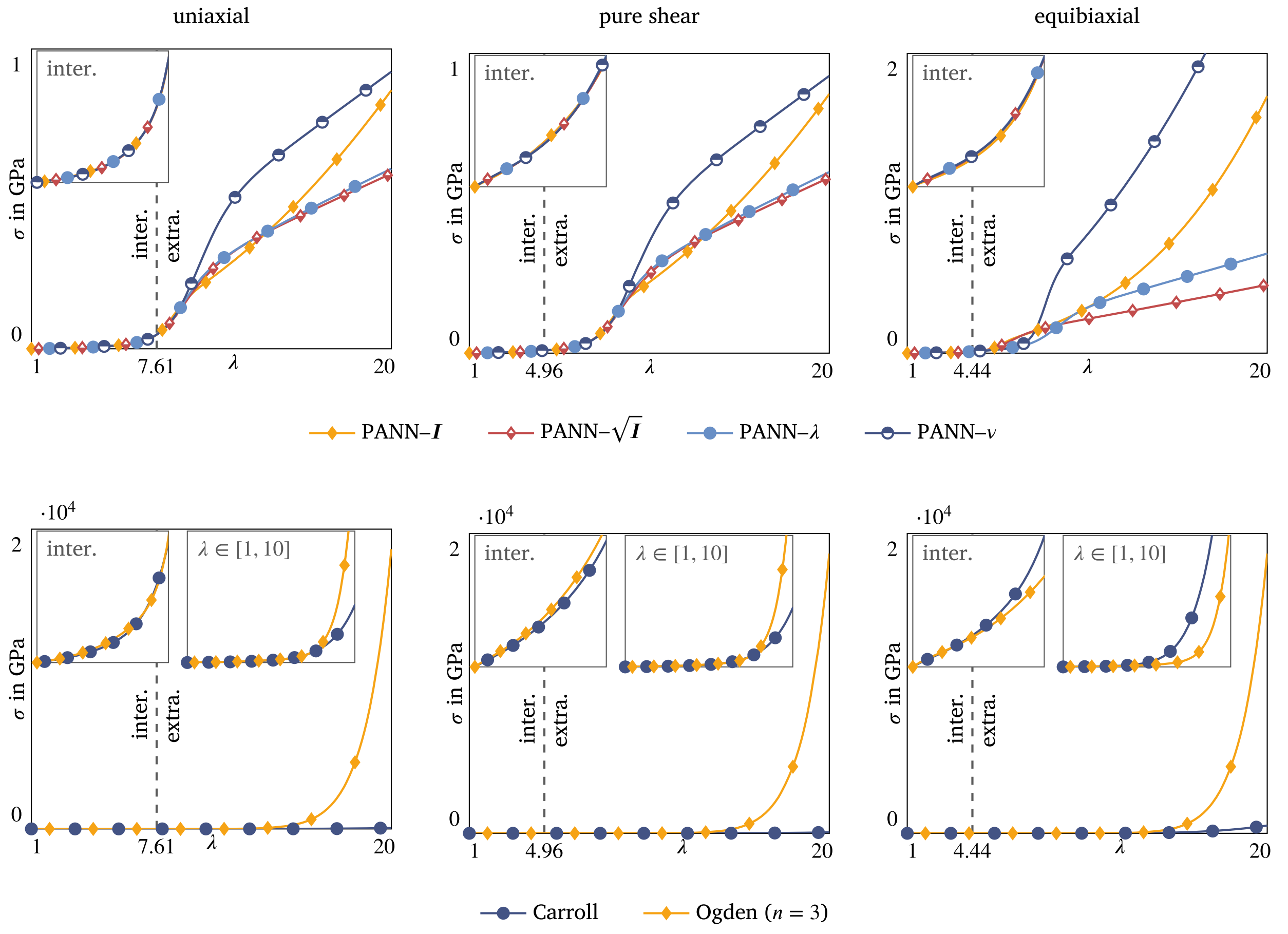

**Fig. 3.** Extrapolation for different constitutive models calibrated to Treloar's data. All deformation modes are used for calibration. The gray vertical line indicates up to which stretch calibration data is available. Top: PANN constitutive models. All PANN models show a very similar performance in interpolation but differ significantly in extrapolation. Bottom: Conventional models with parameters from Carroll (2011) and Steinmann et al. (2012), respectively. The conventional models slightly differ in interpolation, but differ significantly in extrapolation.

To further investigate the extrapolation behavior of the PANN models, we take a closer look at the uniaxial tensile response of the invariant-based PANN–$\boldsymbol{I}/\sqrt{\boldsymbol{I}}$ models, which are again visualized in Fig. 4(a). Similar investigations could be conducted for the PANN–$\lambda/\nu$ models and other load scenarios, which are omitted here for brevity. Following (4.7) and $\boldsymbol{\sigma} = \mathbf{S}_1\mathbf{F}^{\mathrm{T}}$ in case of incompressibility, the Cauchy stress in uniaxial tension of these models is given by

$$\sigma = \left( \underbrace{\frac{\partial \Psi^{\mathrm{NN}}_{\square}}{\partial I^{\square}_1}}_{c_1} \underbrace{\left( \lambda_1 \frac{\partial I^{\square}_1}{\partial \lambda_1} - \lambda_3 \frac{\partial I^{\square}_1}{\partial \lambda_3} \right)}_{d_1} + \underbrace{\frac{\partial \Psi^{\mathrm{NN}}_{\square}}{\partial I^{\square}_2}}_{c_2} \underbrace{\left( \lambda_1 \frac{\partial I^{\square}_2}{\partial \lambda_1} - \lambda_3 \frac{\partial I^{\square}_2}{\partial \lambda_3} \right)}_{d_2} \right)\Bigg|_{\mathbf{F}_{\mathrm{UX}}} \qquad \text{with} \qquad \boldsymbol{I}^{\square} = (I^{\square}_1, I^{\square}_2), \tag{5.3}$$

where the invariants $\boldsymbol{I}^{\square}$ are defined following (4.2). Each of the summands above depends on two functions: $c_1, c_2$ containing the derivatives of the potential w.r.t. the invariants, sometimes referred to as stress coefficients, cf. Kalina et al. (2022, Eq. (7)), and $d_1, d_2$ containing derivatives of the invariants w.r.t. the principal stretches. For both models, the stress coefficients are given by

$$c_l = \frac{\partial \Psi^{\mathrm{NN}}_{\square}}{\partial I^{\square}_l} = \sum_{k=1}^{n} w_k \, \mathrm{SM}(W_{k1} I^{\square}_1 + W_{k2} I^{\square}_2 + b_k) W_{kl} \tag{5.4}$$

with the sigmoid function

$$\mathrm{SM}(x) = \frac{e^x}{1 + e^x} = \frac{\mathrm{d}\,\mathrm{SP}(x)}{\mathrm{d}x}, \tag{5.5}$$

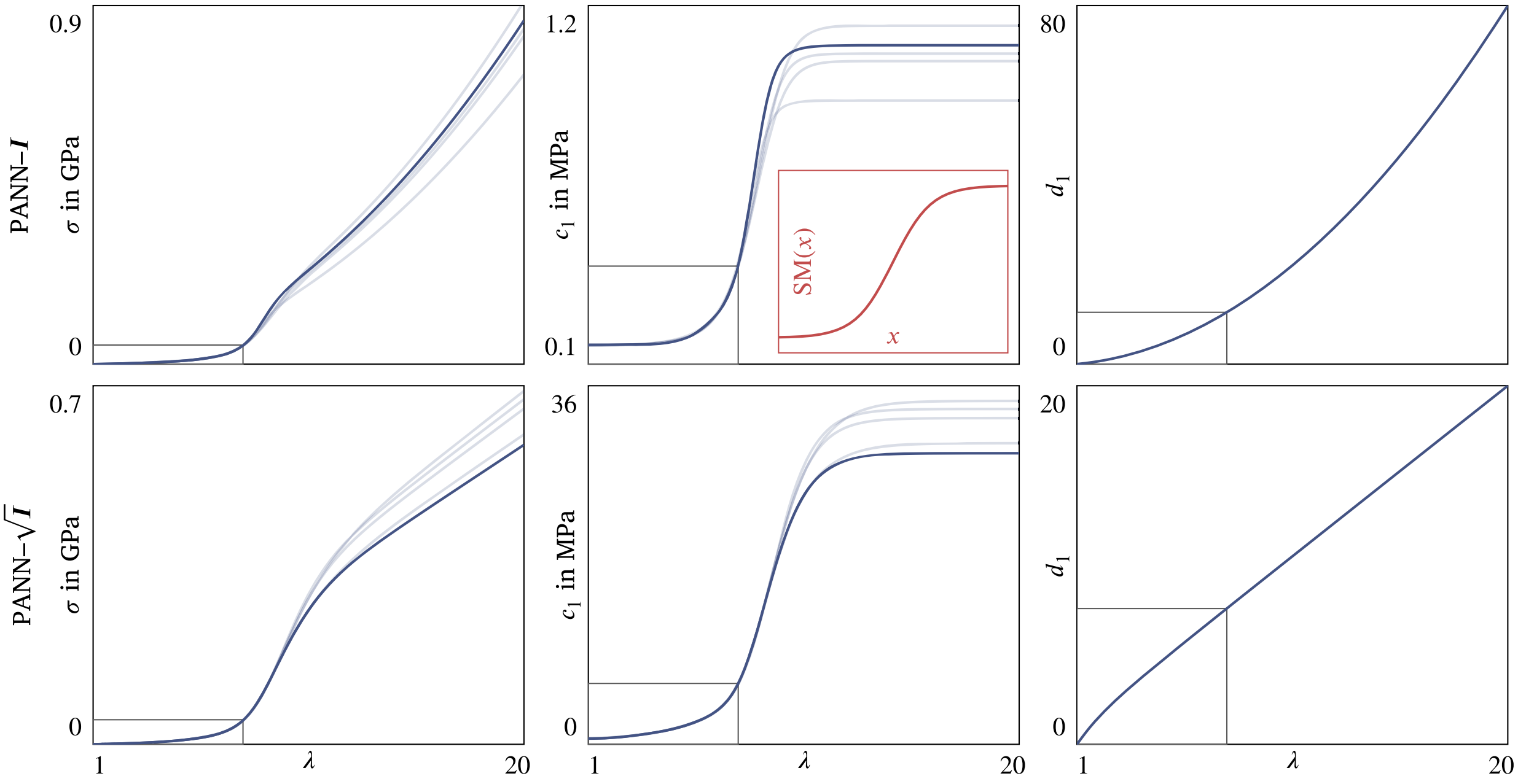

(a) Cauchy stress response

(b) $c_1$ contains derivatives of the potential w.r.t. the invariants

(c) $d_1$ contains derivatives of the invariants w.r.t. the principal stretches

**Fig. 4.** Extrapolation behavior for the PANN–$\boldsymbol{I}/\sqrt{\boldsymbol{I}}$ models calibrated to Treloar's data. The Cauchy stress response depends on $c_{1,2}$ and $d_{1,2}$, cf. (5.3). Five differently calibrated instances of the same model due to random initialization are shown; all but the one with the smallest MSE is depicted transparently. The gray box indicates the calibration regime.

which is visualized in Fig. 4(a,top). The first stress coefficients $c_1$ for the PANN–$\boldsymbol{I}/\sqrt{\boldsymbol{I}}$ models are visualized in Fig. 4(b). For the set of optimal parameters, the stress coefficients resemble sigmoid functions themselves. This entails three distinct regimes: for very small and very large deformations, the stress coefficient is approximately constant, while in the intermediate range it has a steep slope. This transient in between the flat regimes transfers to the stress response of the PANN models, as depicted in Figs. 3 and 4(a). Moreover, the stress response of the PANN–$\boldsymbol{I}/\sqrt{\boldsymbol{I}}$ models is dictated by the derivatives of the different invariants w.r.t. the principal stretches, i.e., $d_1, d_2$ defined in (5.3). As different sets of invariants are employed, their derivatives change accordingly. In our case, we have

$$\text{PANN–}\boldsymbol{I}\text{:}\quad d_1 = 2(\lambda^2 - \lambda^{-1}) \qquad \text{and} \qquad \text{PANN–}\sqrt{\boldsymbol{I}}\text{:}\quad d_1 = \frac{\lambda^2 - \lambda^{-1}}{\sqrt{\lambda^2 + 2\lambda^{-1}}}, \tag{5.6}$$

which are visualized in Fig. 4(c). For large stretches, $d_1$ grows quadratically for the PANN–$\boldsymbol{I}$ model, while increasing only linearly for the PANN–$\sqrt{\boldsymbol{I}}$ model which further explains the pronounced differences in the extrapolation behavior of the different PANN models.

**Remark 5.3.** Our numerical investigations are restricted to a comparatively simple class of PANN models, which are all isotropic, incompressible, and satisfy (virtually) the same constitutive constraints. In that sense, our findings do not necessary apply to other modeling strategies involving neural networks. Nonetheless, within our select group, we could observe that PANN architectures can perform quite differently during training, even though they satisfy the same constitutive constraints. More importantly, however, their extrapolation behavior can vary considerably, even if the goodness of fit is virtually indistinguishable. It seems reasonable to conjecture that these observations apply also to more complex applications of PANNs, e.g., the modeling of compressible, inelastic, and multiphysical material behavior, potentially to an even greater extent.

### *5.4. An open question: Do we need to constrain the curvature of the Cauchy stress for 'idealized elasticity'?*

In light of the previous examples, let us recall one of the main open questions of the theory of material behavior, often referred to as Truesdell's Hauptproblem, cf. Truesdell (1956). If we disregard any inelastic effects such as fatigue, softening, or plasticity, what set of constitutive constraints is required to represent idealized elasticity? Setting aside obvious requirements such as frame indifference, one possible answer appeared to be the combination of polyconvexity (and in turn ellipticity) and TSTS-M, cf. Wollner et al. (2026a). Together, these ensure a monotonically increasing Cauchy stress response for load scenarios where we would expect such a behavior. However, even though in all of our investigations the Cauchy stress response of the PANNs increases monotonically, it does not do so in a manner we might expect. In contrast to the conventional models which become progressively stiffer, see Fig. 3,

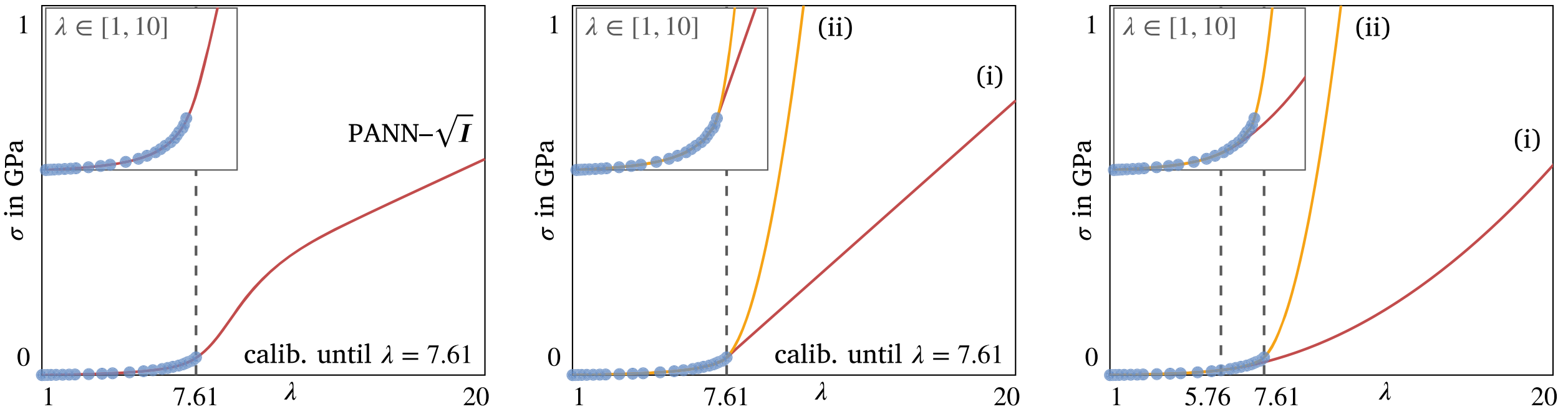

(a) PANN extrapolation with negative curvature

(b) Extrapolation with (i) constant slope and zero curvature and (ii) positive curvature

(c) Extrapolation with positive curvature for a calibration until (i) $\lambda = 5.76$ and (ii) $\lambda = 7.61$

**Fig. 5.** The Cauchy stress of the PANNs is monotonically increasing, partly with negative curvature. For idealized elasticity, one might postulate that the Cauchy stress should be monotonically increasing with a non-negative curvature; this does, however, not resolve the fundamental ambiguity of extrapolation. For illustration, we show Treloar's experimental data in uniaxial tension in blue alongside (hypothetical) model extrapolations in red and yellow.

the PANNs predict a fall-off for $\lambda > 10$, see Fig. 5(a). While the Cauchy stress is still monotonically increasing, its curvature becomes negative. This is in conflict with our understanding of an ideal elastic body, where our intuition suggests that the Cauchy stress should be monotonically increasing with a *non-negative* curvature.

A heuristic choice of such an extrapolation would be to extend the stress curve with a non-negative curvature in addition to a positive slope in basic loading scenarios such as uniaxial tension, see Fig. 5(b). While the simplicity of such choices is appealing, it does not fundamentally resolve the ambiguity entailed by extrapolation, see Fig. 5(c). Such considerations would get even more challenging for general loading scenarios. Apart from the question of how to reasonably formalize a constitutive constraint for the curvature of the Cauchy stress, the inclusion of such a condition in a practical modeling framework would entail further challenges. Lastly, the mathematical notion of ideal elasticity is in conflict with the application of constitutive models to real-world experimental data[28], which never entail solely elastic effects, see also Remark 5.1.

While the combination of polyconvexity and TSTS-M restricts the *slope* of the Cauchy stress, it does not provide sufficient constraints on the *curvature* of the Cauchy stress. Thus, we cautiously conjecture that a constraint on the curvature of the Cauchy stress might have to be included in the set of constitutive conditions to represent idealized elasticity. Then again, this line of thinking can be continued *ad infinitum*. Other authors have rejected the importance of constitutive inequalities altogether, e.g., cf. Ericksen (2013, 36:21).

## 6. Conclusion

In this contribution, we revisited the constitutive constraints of hyperelasticity, specifically in the context of incompressibility. One principal result is that polyconvexity implies Hill's inequality and consequently TSTS-M for a large class of isotropic strain-energy functions. Consequently, we provided several PANN architectures that satisfy these constraints in a sufficient manner. Via calibration to various sets of experimental data, we illustrated the notion that the choice of parametrization has an impact on the approximation power of a PANN, even if constrained to the same set of constitutive requirements. For a guaranteed combination of polyconvexity and TSTS-M, the architecture defined via the conditions by Ball (1976, Thm. 5.2) is both general and powerful. The latter is only outperformed by the architecture based on signed-singular polyconvexity, cf. Geuken et al. (2025, 2026), although here TSTS-M might not be ensured *a priori*. More importantly, even if distinct PANN architectures obey the same constitutive constraints and interpolate with comparable quality, their extrapolation can differ widely. This in turn prompts two questions:

(i) Is our set of constitutive constraints actually sufficient to guarantee physically reasonable extrapolation?
(ii) Which of these roughly equivalent PANN architectures should then be chosen for extrapolation?

Regarding the first question, it might be worthwhile to reconsider the predictive power of pure phenomenology in favor of simpler models that incorporate additional considerations independent from macroscopic experimental data. A classic example might be the strain-energy function by Arruda and Boyce (1993) building on the chain statistics of Kuhn and Grün (1942). What one sacrifices in quantitative accuracy, one gains in control and qualitative understanding. The usefulness of compact closed-form models that allow for simple, but very general conclusions should not be underestimated in a time of computational mechanics.

[28] This has already been acknowledged by Truesdell (1956, p. 7; translation by the authors): 'Doubts have been raised if physical bodies which experience finite, yet purely elastic deformations actually exist, that is, if the ideal elastic body is a mere mathematical curiosity.'

With respect to the second question, one should keep in mind that PANNs are in the end just functions parametrized by a comparatively large number of parameters. Consequently, the manner in which these parameters are inferred from experimental data has an influence on the final calibrated PANN and its predictions. The delicate interplay between number of model parameters and number of data points has been recently discussed in Wollner et al. (2026b) and seems especially relevant for the training of PANNs. There, we should expect a whole range of parameter combinations to give reasonable fits, while potentially extrapolating very differently, cf. Linka et al. (2025). While the minimization of a non-convex loss function can be repeated and analyzed for different initialization, the choice of the loss function itself essentially becomes a modeling choice.

Regarding the choice of PANN architecture, one can obviously argue from a numerical perspective. For instance, different architectures entail different computational costs, as alluded to in Remark 4.1. Another aspect might be numerical stability and robustness in finite-element applications. However, from a purely physical perspective, the honest answer appears to be that within a set of models, that each satisfy the same constitutive constraints *a priori*, it is not possible to determine which one should be preferred. One key takeaway from this contribution is that a physics-augmented ansatz with high approximation power does not automatically yield a material model with definite predictive power. Finally, we emphasize that the points we have raised here do not only apply to the constitutive modeling of incompressible hyperelasticity, but extend to more complex applications of PANNs, most likely in an exacerbated fashion.

## CRediT authorship contribution statement

**Maximilian P. Wollner:** Writing – review & editing, Writing – original draft, Methodology, Investigation, Formal analysis, Conceptualization; **Dominik K. Klein:** Writing – review & editing, Writing – original draft, Visualization, Software, Methodology, Investigation, Formal analysis, Conceptualization; **Herbert Baaser:** Writing – review & editing, Methodology, Conceptualization; **Gerhard A. Holzapfel:** Writing – review & editing, Supervision, Funding acquisition; **Patrizio Neff:** Writing – review & editing, Writing – original draft, Supervision, Methodology, Conceptualization.

## AI declaration

Maximilian P. Wollner used an enterprise version of GPT-5 by OpenAI to identify typographical and grammatical errors. Dominik K. Klein used GPT-5.5 by OpenAI for select linguistic refinements.

## Declaration of competing interest

The authors declare that they have no known competing financial interests or personal relationships that could have appeared to influence the work reported in this paper.

## Acknowledgments

Dominik K. Klein acknowledges the financial support provided by the Deutsche Forschungsgemeinschaft (DFG, German Research Foundation, project number 492770117) and the Graduate School of Computational Engineering at TU Darmstadt.

## Data availability

Only previously published experimental data was used for the research described in the article. For the original data, readers are referred to the corresponding publications.

## Appendix A. Derivation of the necessary and sufficient conditions for Hill's inequality in case of incompressibility

For the sake of completeness and traceability, we present a derivation of the necessary and sufficient conditions for Hill's inequality (2.8) in case of incompressibility as well as the implication (2.14).

### A.1. A particular representation of Hill's inequality

In case of incompressibility, we have $\det \mathbf{F} = 1$ and $\operatorname{tr}\mathbf{D} = 0$. Furthermore, an additional Lagrange parameter $\tilde{p}$ enters the elastic stress response which importantly leaves Hill's inequality (2.9) unchanged, since

$$\left\langle \frac{\mathrm{D}^{\mathrm{ZJ}}(\tilde{p}\,\mathbb{1})}{\mathrm{D}t}, \mathbf{D} \right\rangle = \left\langle \dot{\tilde{p}}\,\mathbb{1} + \tilde{p}\,\mathbf{W} - \tilde{p}\,\mathbf{W}, \mathbf{D} \right\rangle = \dot{\tilde{p}}\operatorname{tr}\mathbf{D}, \tag{A.1}$$

such that

$$\left\langle \frac{\mathrm{D}^{\mathrm{ZJ}}(\boldsymbol{\tau} + \tilde{p}\,\mathbb{1})}{\mathrm{D}t}, \mathbf{D} \right\rangle = \left\langle \frac{\mathrm{D}^{\mathrm{ZJ}}\boldsymbol{\tau}}{\mathrm{D}t}, \mathbf{D} \right\rangle > 0 \qquad \forall \operatorname{tr}\mathbf{D} = 0. \tag{A.2}$$

The rate constraint above has been analyzed by Hill (1968, 1970) through the use of the 'material' and 'spatial strain ellipsoid'; a similar approach was used in Leblond (1992) for (2.3). Here, we express (A.2) slightly differently. We may write the deformation gradient $\mathbf{F}$ by virtue of its singular value decomposition as

$$\mathbf{F} = \sum_{\mathfrak{a}=1}^{3} \lambda_{\mathfrak{a}}\, \boldsymbol{n}_{\mathfrak{a}} \otimes \boldsymbol{N}_{\mathfrak{a}}, \tag{A.3}$$

where $\lambda_{\mathfrak{a}}$ denote the principal stretches and $\boldsymbol{n}_{\mathfrak{a}}$, $\boldsymbol{N}_{\mathfrak{a}}$ the associated material and spatial principal directions, respectively.[29] Remember that

$$\langle \boldsymbol{n}_{\mathfrak{a}}, \boldsymbol{n}_{\mathfrak{b}} \rangle = \delta_{\mathfrak{a}\mathfrak{b}} \qquad \Longrightarrow \qquad \langle \dot{\boldsymbol{n}}_{\mathfrak{a}}, \boldsymbol{n}_{\mathfrak{b}} \rangle = -\langle \boldsymbol{n}_{\mathfrak{a}}, \dot{\boldsymbol{n}}_{\mathfrak{b}} \rangle. \tag{A.4}$$

Furthermore,

$$\sum_{\mathfrak{a}=1}^{3} \boldsymbol{n}_{\mathfrak{a}} \otimes \boldsymbol{n}_{\mathfrak{a}} = \mathbb{1} \qquad \Longrightarrow \qquad \sum_{\mathfrak{a}=1}^{3} \dot{\boldsymbol{n}}_{\mathfrak{a}} \otimes \boldsymbol{n}_{\mathfrak{a}} = -\sum_{\mathfrak{a}=1}^{3} \boldsymbol{n}_{\mathfrak{a}} \otimes \dot{\boldsymbol{n}}_{\mathfrak{a}}. \tag{A.5}$$

Both identities hold analogously for $\boldsymbol{N}_{\mathfrak{a}}$. With this, we have

$$\begin{aligned} \mathbf{L} = \dot{\mathbf{F}}\mathbf{F}^{-1} &= \sum_{\mathfrak{a}}^{3} \Big( \dot{\lambda}_{\mathfrak{a}}\, \boldsymbol{n}_{\mathfrak{a}} \otimes \boldsymbol{N}_{\mathfrak{a}} + \lambda_{\mathfrak{a}}\, \dot{\boldsymbol{n}}_{\mathfrak{a}} \otimes \boldsymbol{N}_{\mathfrak{a}} + \lambda_{\mathfrak{a}}\, \boldsymbol{n}_{\mathfrak{a}} \otimes \dot{\boldsymbol{N}}_{\mathfrak{a}} \Big) \sum_{\mathfrak{b}=1}^{3} \lambda_{\mathfrak{b}}^{-1} \boldsymbol{N}_{\mathfrak{b}} \otimes \boldsymbol{n}_{\mathfrak{b}} \\ &= \sum_{\mathfrak{a}=1} \frac{\dot{\lambda}_{\mathfrak{a}}}{\lambda_{\mathfrak{a}}}\, \boldsymbol{n}_{\mathfrak{a}} \otimes \boldsymbol{n}_{\mathfrak{a}} + \sum_{\mathfrak{a}=1}^{3} \dot{\boldsymbol{n}}_{\mathfrak{a}} \otimes \boldsymbol{n}_{\mathfrak{a}} + \sum_{\mathfrak{a}=1}^{3} \sum_{\mathfrak{b}=1}^{3} \frac{\lambda_{\mathfrak{a}}}{\lambda_{\mathfrak{b}}}\, \Omega_{\mathfrak{a}\mathfrak{b}}\, \boldsymbol{n}_{\mathfrak{a}} \otimes \boldsymbol{n}_{\mathfrak{b}}, \end{aligned} \tag{A.6}$$

where $\Omega_{\mathfrak{a}\mathfrak{b}} = \langle \dot{\boldsymbol{N}}_{\mathfrak{a}}, \boldsymbol{N}_{\mathfrak{b}} \rangle = -\Omega_{\mathfrak{b}\mathfrak{a}}$. It follows that

$$\begin{aligned} \mathbf{D} = \frac{1}{2}(\mathbf{L} + \mathbf{L}^{\mathrm{T}}) &= \sum_{\mathfrak{a}=1}^{3} \frac{\dot{\lambda}_{\mathfrak{a}}}{\lambda_{\mathfrak{a}}}\, \boldsymbol{n}_{\mathfrak{a}} \otimes \boldsymbol{n}_{\mathfrak{a}} + \frac{1}{2} \sum_{\mathfrak{a}=1}^{3} \sum_{\mathfrak{b}=1}^{3} \frac{\lambda_{\mathfrak{a}}}{\lambda_{\mathfrak{b}}}\, \Omega_{\mathfrak{a}\mathfrak{b}} \big( \boldsymbol{n}_{\mathfrak{a}} \otimes \boldsymbol{n}_{\mathfrak{b}} + \boldsymbol{n}_{\mathfrak{b}} \otimes \boldsymbol{n}_{\mathfrak{a}} \big) \\ &= \sum_{\mathfrak{a}=1}^{3} \frac{\dot{\lambda}_{\mathfrak{a}}}{\lambda_{\mathfrak{a}}}\, \boldsymbol{n}_{\mathfrak{a}} \otimes \boldsymbol{n}_{\mathfrak{a}} + \frac{1}{2} \sum_{\mathfrak{a}=1}^{3} \sum_{\mathfrak{b}<\mathfrak{a}} \Big( \frac{\lambda_{\mathfrak{a}}}{\lambda_{\mathfrak{b}}} - \frac{\lambda_{\mathfrak{b}}}{\lambda_{\mathfrak{a}}} \Big) \Omega_{\mathfrak{a}\mathfrak{b}} \big( \boldsymbol{n}_{\mathfrak{a}} \otimes \boldsymbol{n}_{\mathfrak{b}} + \boldsymbol{n}_{\mathfrak{b}} \otimes \boldsymbol{n}_{\mathfrak{a}} \big), \end{aligned} \tag{A.7}$$

and

$$\begin{aligned} \mathbf{W} = \frac{1}{2}(\mathbf{L} - \mathbf{L}^{\mathrm{T}}) &= \sum_{\mathfrak{a}=1}^{3} \dot{\boldsymbol{n}}_{\mathfrak{a}} \otimes \boldsymbol{n}_{\mathfrak{a}} + \frac{1}{2} \sum_{\mathfrak{a}=1}^{3} \sum_{\mathfrak{b}=1}^{3} \frac{\lambda_{\mathfrak{a}}}{\lambda_{\mathfrak{b}}}\, \Omega_{\mathfrak{a}\mathfrak{b}} \big( \boldsymbol{n}_{\mathfrak{a}} \otimes \boldsymbol{n}_{\mathfrak{b}} - \boldsymbol{n}_{\mathfrak{b}} \otimes \boldsymbol{n}_{\mathfrak{a}} \big) \\ &= -\sum_{\mathfrak{a}=1}^{3} \boldsymbol{n}_{\mathfrak{a}} \otimes \dot{\boldsymbol{n}}_{\mathfrak{a}} + \frac{1}{2} \sum_{\mathfrak{a}=1}^{3} \sum_{\mathfrak{b}<\mathfrak{a}} \Big( \frac{\lambda_{\mathfrak{a}}}{\lambda_{\mathfrak{b}}} + \frac{\lambda_{\mathfrak{b}}}{\lambda_{\mathfrak{a}}} \Big) \Omega_{\mathfrak{a}\mathfrak{b}} \big( \boldsymbol{n}_{\mathfrak{a}} \otimes \boldsymbol{n}_{\mathfrak{b}} - \boldsymbol{n}_{\mathfrak{b}} \otimes \boldsymbol{n}_{\mathfrak{a}} \big). \end{aligned} \tag{A.8}$$

In case of isotropy, we may express the Kirchhoff stress $\boldsymbol{\tau}$ in its spectral decomposition with

$$\boldsymbol{\tau} = \sum_{\mathfrak{a}=1}^{3} \tau_{\mathfrak{a}}\, \boldsymbol{n}_{\mathfrak{a}} \otimes \boldsymbol{n}_{\mathfrak{a}}, \tag{A.9}$$

where $\tau_{\mathfrak{a}}$ denotes the principal Kirchhoff stresses. For the Zaremba–Jaumann rate we need to evaluate

$$\begin{aligned} \boldsymbol{\tau}\,\mathbf{W} &= \sum_{\mathfrak{c}=1}^{3} \tau_{\mathfrak{c}}\, \boldsymbol{n}_{\mathfrak{c}} \otimes \boldsymbol{n}_{\mathfrak{c}} \bigg( -\sum_{\mathfrak{a}=1}^{3} \boldsymbol{n}_{\mathfrak{a}} \otimes \dot{\boldsymbol{n}}_{\mathfrak{a}} + \frac{1}{2} \sum_{\mathfrak{a}=1}^{3} \sum_{\mathfrak{b}<\mathfrak{a}} \Big( \frac{\lambda_{\mathfrak{a}}}{\lambda_{\mathfrak{b}}} + \frac{\lambda_{\mathfrak{b}}}{\lambda_{\mathfrak{a}}} \Big) \Omega_{\mathfrak{a}\mathfrak{b}} \big( \boldsymbol{n}_{\mathfrak{a}} \otimes \boldsymbol{n}_{\mathfrak{b}} - \boldsymbol{n}_{\mathfrak{b}} \otimes \boldsymbol{n}_{\mathfrak{a}} \big) \bigg) \\ &= -\sum_{\mathfrak{a}=1}^{3} \tau_{\mathfrak{a}}\, \boldsymbol{n}_{\mathfrak{a}} \otimes \dot{\boldsymbol{n}}_{\mathfrak{a}} + \frac{1}{2} \sum_{\mathfrak{a}=1}^{3} \sum_{\mathfrak{b}<\mathfrak{a}} \Big( \frac{\lambda_{\mathfrak{a}}}{\lambda_{\mathfrak{b}}} + \frac{\lambda_{\mathfrak{b}}}{\lambda_{\mathfrak{a}}} \Big) \Omega_{\mathfrak{a}\mathfrak{b}} \big( \tau_{\mathfrak{a}}\, \boldsymbol{n}_{\mathfrak{a}} \otimes \boldsymbol{n}_{\mathfrak{b}} - \tau_{\mathfrak{b}} \boldsymbol{n}_{\mathfrak{b}} \otimes \boldsymbol{n}_{\mathfrak{a}} \big) \end{aligned} \tag{A.10}$$

and

$$\begin{aligned} &\boldsymbol{\tau}\,\mathbf{W} - \mathbf{W}\,\boldsymbol{\tau} = \boldsymbol{\tau}\,\mathbf{W} + (\boldsymbol{\tau}\,\mathbf{W})^{\mathrm{T}} \\ &\quad = -\sum_{\mathfrak{a}=1}^{3} \tau_{\mathfrak{a}} \big( \boldsymbol{n}_{\mathfrak{a}} \otimes \dot{\boldsymbol{n}}_{\mathfrak{a}} + \dot{\boldsymbol{n}}_{\mathfrak{a}} \otimes \boldsymbol{n}_{\mathfrak{a}} \big) + \frac{1}{2} \sum_{\mathfrak{a}=1}^{3} \sum_{\mathfrak{b}<\mathfrak{a}} \Big( \frac{\lambda_{\mathfrak{a}}}{\lambda_{\mathfrak{b}}} + \frac{\lambda_{\mathfrak{b}}}{\lambda_{\mathfrak{a}}} \Big) (\tau_{\mathfrak{a}} - \tau_{\mathfrak{b}})\, \Omega_{\mathfrak{a}\mathfrak{b}} \big( \boldsymbol{n}_{\mathfrak{a}} \otimes \boldsymbol{n}_{\mathfrak{b}} + \boldsymbol{n}_{\mathfrak{b}} \otimes \boldsymbol{n}_{\mathfrak{a}} \big). \end{aligned} \tag{A.11}$$

Further,

$$\dot{\boldsymbol{\tau}} + \boldsymbol{\tau}\,\mathbf{W} - \mathbf{W}\,\boldsymbol{\tau} = \sum_{\mathfrak{a}=1}^{3} \dot{\tau}_{\mathfrak{a}}\, \boldsymbol{n}_{\mathfrak{a}} \otimes \boldsymbol{n}_{\mathfrak{a}} \qquad + \frac{1}{2} \sum_{\mathfrak{a}=1}^{3} \sum_{\mathfrak{b}<\mathfrak{a}} \Big( \frac{\lambda_{\mathfrak{a}}}{\lambda_{\mathfrak{b}}} + \frac{\lambda_{\mathfrak{b}}}{\lambda_{\mathfrak{a}}} \Big) (\tau_{\mathfrak{a}} - \tau_{\mathfrak{b}})\, \Omega_{\mathfrak{a}\mathfrak{b}} \big( \boldsymbol{n}_{\mathfrak{a}} \otimes \boldsymbol{n}_{\mathfrak{b}} + \boldsymbol{n}_{\mathfrak{b}} \otimes \boldsymbol{n}_{\mathfrak{a}} \big) \tag{A.12}$$

and finally

$$\big\langle \dot{\boldsymbol{\tau}} + \boldsymbol{\tau}\,\mathbf{W} - \mathbf{W}\,\boldsymbol{\tau}, \mathbf{D} \big\rangle = \sum_{\mathfrak{a}=1}^{3} \dot{\tau}_{\mathfrak{a}} \frac{\dot{\lambda}_{\mathfrak{a}}}{\lambda_{\mathfrak{a}}} + \sum_{\mathfrak{a}=1}^{3} \sum_{\mathfrak{b}<\mathfrak{a}} \frac{(\lambda_{\mathfrak{a}}^4 - \lambda_{\mathfrak{b}}^4)(\tau_{\mathfrak{a}} - \tau_{\mathfrak{b}})}{\lambda_{\mathfrak{a}}^2 \lambda_{\mathfrak{b}}^2}\, \Omega_{\mathfrak{a}\mathfrak{b}}^2 > 0 \qquad \forall\ \mathrm{tr}\,\mathbf{D} = \sum_{\mathfrak{a}=1}^{3} \frac{\dot{\lambda}_{\mathfrak{a}}}{\lambda_{\mathfrak{a}}} = 0. \tag{A.13}$$

[29] The singular value decomposition in (A.3) need not be unique. For our purposes, however, any such representation is enough.

### A.2. Necessary and sufficient conditions for the weak form of Hill's inequality

In case of isotropic hyperelasticity, we have a strain-energy function $\widehat{W}(\log\lambda_1,\log\lambda_2,\log\lambda_3)$ such that

$$\tau_{\mathfrak{a}}=\frac{\partial\widehat{W}}{\partial\log\lambda_{\mathfrak{a}}}\qquad\Longrightarrow\qquad\dot{\tau}_{\mathfrak{a}}=\frac{\partial^2\widehat{W}}{\partial\log\lambda_{\mathfrak{a}}\,\partial\log\lambda_{\mathfrak{b}}}\,\frac{\dot{\lambda}_{\mathfrak{b}}}{\lambda_{\mathfrak{b}}}. \tag{A.14}$$

For an irrotational motion $\Omega_{\mathfrak{ab}}=0$ in (A.13), we then have the necessary condition

$$\sum_{\mathfrak{a}=1}^{3}\sum_{\mathfrak{b}=1}^{3}\frac{\dot{\lambda}_{\mathfrak{a}}}{\lambda_{\mathfrak{a}}}\,\frac{\partial^2\widehat{W}}{\partial\log\lambda_{\mathfrak{a}}\,\partial\log\lambda_{\mathfrak{b}}}\,\frac{\dot{\lambda}_{\mathfrak{b}}}{\lambda_{\mathfrak{b}}}>0\qquad\forall\sum_{\mathfrak{a}=1}^{3}\frac{\dot{\lambda}_{\mathfrak{a}}}{\lambda_{\mathfrak{a}}}=0. \tag{A.15}$$

The expression above is clearly positive, if $\widehat{W}$ is strictly convex, but only sufficiently so, because of the constraint $\operatorname{tr}\mathbf{D}=0$. In fact, if we work with the projected Hessian[30] instead to incorporate the constraint, we find that only

$$\begin{bmatrix}\widehat{W}_{11}-2\widehat{W}_{13}+\widehat{W}_{33} & \widehat{W}_{12}-\widehat{W}_{13}-\widehat{W}_{23}+\widehat{W}_{33}\\ \text{sym.} & \widehat{W}_{22}-2\widehat{W}_{23}+\widehat{W}_{33}\end{bmatrix}\qquad\text{with}\qquad\widehat{W}_{\mathfrak{ab}}=\frac{\partial^2\widehat{W}}{\partial\log\lambda_{\mathfrak{a}}\,\partial\log\lambda_{\mathfrak{b}}} \tag{A.16}$$

must be positive-definite for (A.15) to hold, cf. Baaser (2026a, App. D.2). This in turn is equivalent to requiring strict convexity of

$$\widehat{W}^{\text{inc}}_{\text{red}}(\log\lambda_1,\log\lambda_2):=\widehat{W}\big(\log\lambda_1,\log\lambda_2,-\log\lambda_1-\log\lambda_2\big), \tag{A.17}$$

cf. Baaser (2026a, App. D.3).

Returning back to (A.13), we now show that

$$\big(\lambda_{\mathfrak{a}}^4-\lambda_{\mathfrak{b}}^4\big)(\tau_{\mathfrak{a}}-\tau_{\mathfrak{b}})>0 \tag{A.18}$$

is implied by strict convexity of $\widehat{W}^{\text{inc}}_{\text{red}}$ for all non-trivial deformation states. First, notice that

$$\big(\lambda_{\mathfrak{a}}^4-\lambda_{\mathfrak{b}}^4\big)(\tau_{\mathfrak{a}}-\tau_{\mathfrak{b}})>0\quad\forall\,\lambda_{\mathfrak{a}}\neq\lambda_{\mathfrak{b}}\qquad\Longleftrightarrow\qquad(\tau_{\mathfrak{a}}-\tau_{\mathfrak{b}})\big(\log\lambda_{\mathfrak{a}}-\log\lambda_{\mathfrak{b}}\big)>0\quad\forall\,\lambda_{\mathfrak{a}}\neq\lambda_{\mathfrak{b}}. \tag{A.19}$$

Second, with (A.14) and (A.17), the above constraint is equivalent to

$$\frac{\partial\widehat{W}^{\text{inc}}_{\text{red}}}{\partial\log\lambda_1}(\log\lambda_1-\log\lambda_3)>0,\qquad\forall\,\lambda_1\neq\lambda_3 \tag{A.20}$$

$$\frac{\partial\widehat{W}^{\text{inc}}_{\text{red}}}{\partial\log\lambda_2}(\log\lambda_2-\log\lambda_3)>0,\qquad\forall\,\lambda_2\neq\lambda_3 \tag{A.21}$$

$$\left(\frac{\partial\widehat{W}^{\text{inc}}_{\text{red}}}{\partial\log\lambda_1}-\frac{\partial\widehat{W}^{\text{inc}}_{\text{red}}}{\partial\log\lambda_2}\right)\big(\log\lambda_1-\log\lambda_2\big)>0,\qquad\forall\,\lambda_1\neq\lambda_2. \tag{A.22}$$

The last condition follows immediately from the strict convexity and permutation-invariance of $\widehat{W}^{\text{inc}}_{\text{red}}$. The other two should also hold by symmetry. We can make this explicit, though. Observe that

$$\begin{aligned}\widehat{W}^{\text{inc}}_{\text{red}}(\log\lambda_1,\log\lambda_2)&=\widehat{W}(\log\lambda_1,\log\lambda_2,-\log\lambda_1-\log\lambda_2)\\&=\widehat{W}(-\log\lambda_1-\log\lambda_2,\log\lambda_2,\log\lambda_1)\\&=\widehat{W}^{\text{inc}}_{\text{red}}(-\log\lambda_1-\log\lambda_2,\log\lambda_2).\end{aligned} \tag{A.23}$$

Hence, $\widehat{W}^{\text{inc}}_{\text{red}}$ is symmetric such that

$$\widehat{W}^{\text{inc}}_{\text{red}}(\log\lambda_1,\log\lambda_2)=\widehat{W}^{\text{inc}}_{\text{red}}(-\log\lambda_1-\log\lambda_2,\log\lambda_2) \tag{A.24}$$

with a line of symmetry at $2\log\lambda_1+\log\lambda_2=0$ or equivalently $\lambda_1=\lambda_3$. Since $\widehat{W}^{\text{inc}}_{\text{red}}$ is strictly convex in $\log\lambda_1$ for each fixed $\lambda_2$ with a global minimum at the line of symmetry $\lambda_1=\lambda_3$, it follows that

$$\frac{\partial\widehat{W}^{\text{inc}}_{\text{red}}}{\partial\log\lambda_1}(\log\lambda_1-\log\lambda_3)>0\qquad\forall\lambda_1\neq\lambda_3. \tag{A.25}$$

The argument can be repeated analogously for (A.21) by swapping $\lambda_1$ for $\lambda_2$ resulting in

$$\frac{\partial\widehat{W}^{\text{inc}}_{\text{red}}}{\partial\log\lambda_2}(\log\lambda_2-\log\lambda_3)>0\qquad\forall\lambda_2\neq\lambda_3. \tag{A.26}$$

At this point, we have almost established that strict convexity of $\widehat{W}^{\text{inc}}_{\text{red}}$ is necessary and sufficient for (A.13) to hold. We only need to discuss some fringe cases for which Hill's inequality could vanish. For (A.13) to be zero, we need $\dot{\lambda}_{\mathfrak{a}}=0$ and

$$\lambda_{\mathfrak{a}}=\lambda_{\mathfrak{b}}\qquad\text{or}\qquad\Omega_{\mathfrak{ab}}=0\qquad\forall\mathfrak{a}\neq\mathfrak{b}. \tag{A.27}$$

But in this case, we always have $\mathbf{D}=\mathbf{0}$ by virtue of (A.7) and the associated motion is deemed trivial by definition.

[30] The definiteness in (A.16) is closely related to a stability criterion in the commercial finite element solver Abaqus and Ansys, cf. Baaser (2026a).

### A.3. Equivalence with finite differences

Satisfaction of Hill's inequality (A.2) in case of incompressibility also implies the global constraint

$$\left\langle \widehat{\boldsymbol{\tau}}(\log \mathbf{V}_2) - \widehat{\boldsymbol{\tau}}(\log \mathbf{V}_1), \log \mathbf{V}_2 - \log \mathbf{V}_1 \right\rangle > 0 \qquad \forall\, \mathbf{V}_1 \neq \mathbf{V}_2 \quad \forall\, \det \mathbf{V}_1 = \det \mathbf{V}_2 = 1. \tag{A.28}$$

From here, we abbreviate $\boldsymbol{\tau}_1 = \widehat{\boldsymbol{\tau}}(\log \mathbf{V}_1)$ and $\boldsymbol{\tau}_2 = \widehat{\boldsymbol{\tau}}(\log \mathbf{V}_2)$. First, notice that any additional Lagrange parameter $\tilde{p}$ vanishes since

$$\begin{aligned}\left\langle \boldsymbol{\tau}_2 - \tilde{p}_2 \mathbb{1} - \boldsymbol{\tau}_1 + \tilde{p}_1 \mathbb{1}, \log \mathbf{V}_2 - \log \mathbf{V}_1 \right\rangle &= \left\langle \boldsymbol{\tau}_2 - \boldsymbol{\tau}_1, \log \mathbf{V}_2 - \log \mathbf{V}_1 \right\rangle - (\tilde{p}_2 - \tilde{p}_1)\big(\operatorname{tr} \log \mathbf{V}_2 - \operatorname{tr} \log \mathbf{V}_1\big) \\ &= \left\langle \boldsymbol{\tau}_2 - \boldsymbol{\tau}_1, \log \mathbf{V}_2 - \log \mathbf{V}_1 \right\rangle\end{aligned} \tag{A.29}$$

with $\operatorname{tr} \log \mathbf{V} = \log \det \mathbf{V} = 0$. Then from (A.2), it follows that

$$\begin{aligned}\left\langle \boldsymbol{\tau}_2 - \boldsymbol{\tau}_1, \log \mathbf{V}_2 - \log \mathbf{V}_1 \right\rangle &= \left\langle \int_0^1 \frac{\mathrm{d}}{\mathrm{d}t}\Big(\mathrm{D}_{\log \mathbf{V}} \widehat{W}\big(t \log \mathbf{V}_2 + (1-t) \log \mathbf{V}_1\big)\Big)\, \mathrm{d}t, \log \mathbf{V}_2 - \log \mathbf{V}_1 \right\rangle \\ &= \int_0^1 \left\langle \mathrm{D}^2_{\log \mathbf{V}} \widehat{W}\big(t \log \mathbf{V}_2 + (1-t) \log \mathbf{V}_1\big).\big(\log \mathbf{V}_2 - \log \mathbf{V}_1\big), \log \mathbf{V}_2 - \log \mathbf{V}_1 \right\rangle \mathrm{d}t,\end{aligned} \tag{A.30}$$

cf. Neff et al. (2015, Rem. 4.1). Since $\operatorname{tr} \log \mathbf{V} = 0$ by definition, we have

$$\operatorname{tr}\big(t \log \mathbf{V}_2 + (1-t) \log \mathbf{V}_1\big) = t \operatorname{tr} \log \mathbf{V}_2 + (1-t) \operatorname{tr} \log \mathbf{V}_1 = 0 \qquad \forall\, t, \tag{A.31}$$

i.e., all linear combinations of isochoric Hencky strains are themselves isochoric. Consequently, the integrand in (A.30) is always positive if $\widehat{W}^{\text{inc}}_{\text{red}}$ from (A.17) is strictly convex, as discussed following (A.15). The monotonicity result (A.28) follows accordingly.

The converse is also true, i.e., (A.28) implies convexity of $\widehat{W}^{\text{inc}}_{\text{red}}$. To see this, take $\log \mathbf{V}_2 = \log \mathbf{V} + \delta\, \mathbf{H}$ and $\mathbf{V}_1 = \mathbf{V}$, where $\delta \in \mathbb{R}$ and $\operatorname{tr} \mathbf{H} = 0$ to satisfy incompressibility. Consequently, the inequality (A.28) becomes

$$\left\langle \widehat{\boldsymbol{\tau}}(\log \mathbf{V} + \delta\, \mathbf{H}) - \widehat{\boldsymbol{\tau}}(\log \mathbf{V}), \delta\, \mathbf{H} \right\rangle > 0. \tag{A.32}$$

Multiplying both sides by $\delta^{-2}$ and taking the limit $\delta \to 0$, we read

$$\lim_{\delta \to 0} \left\langle \frac{\boldsymbol{\tau}(\log \mathbf{V} + \delta\, \mathbf{H}) - \widehat{\boldsymbol{\tau}}(\log \mathbf{V})}{\delta}, \mathbf{H} \right\rangle = \left\langle \mathrm{D}_{\log \mathbf{V}} \widehat{\boldsymbol{\tau}}.\mathbf{H}, \mathbf{H} \right\rangle = \left\langle \mathrm{D}^2_{\log \mathbf{V}} \widehat{W}.\mathbf{H}, \mathbf{H} \right\rangle > 0 \quad \forall\, \operatorname{tr} \mathbf{H} = 0. \tag{A.33}$$

We recover convexity of $\widehat{W}$ in $\log \mathbf{V}$ for all $\operatorname{tr} \mathbf{H} = 0$, i.e., by definition the convexity of $\widehat{W}^{\text{inc}}_{\text{red}}$.